\documentclass[a4paper,fleqn]{cas-dc}
\usepackage[authoryear]{natbib}
\usepackage{enumitem}
\usepackage{float}
\usepackage{ragged2e} 
\usepackage{amsmath}
\usepackage{multicol} 
\usepackage{caption}
\usepackage{float}
\usepackage{booktabs}
\usepackage{graphicx}
\usepackage{subcaption}
\usepackage{hyperref}

\definecolor{rqbg}{HTML}{edf9ff}
\definecolor{rqheader}{HTML}{267a9b}  

\usepackage[dvipsnames]{xcolor}
\usepackage[most]{tcolorbox}
\usepackage[table]{xcolor}

\definecolor{A}{RGB}{255,248,220} 
\definecolor{B}{RGB}{255,235,170} 
\definecolor{avggray}{RGB}{235,235,235} 
\definecolor{green}{RGB}{190,225,190} 
\usepackage{mdframed}

\definecolor{mypurple}{HTML}{d4ceff}

\def\tsc#1{\csdef{#1}{\textsc{\lowercase{#1}}\xspace}}
\tsc{WGM}
\tsc{QE}
\tsc{EP}
\tsc{PMS}
\tsc{BEC}
\tsc{DE}

\begin{document}
\let\WriteBookmarks\relax
\def\floatpagepagefraction{1}
\def\textpagefraction{.001}
\shortauthors{Faria et~al.}
\shorttitle{Explainable Convolutional Neural Networks for Retinal Fundus Classification and Cutting-Edge Segmentation Models for Retinal Blood Vessels from Fundus Images}

\title[mode=title]{Explainable Convolutional Neural Networks for Retinal Fundus Classification and Cutting-Edge Segmentation Models for Retinal Blood Vessels from Fundus Images}

\author[1]{Fatema Tuj Johora Faria} [orcid=0009-0005-1684-8211]

\address[1]{Department of Computer Science and Engineering, Ahsanullah University of Science and Technology, Dhaka, Bangladesh}
\ead{fatema.faria142@gmail.com}

\author[1]{Mukaffi Bin Moin} [orcid=0009-0001-9634-5809]
\cormark[1]
\ead{mukaffi28@gmail.com}
\author[1]{Pronay Debnath} [orcid=0009-0006-3900-5345]
\ead{pronaydebnath99@gmail.com}

\author[1]{Asif Iftekher Fahim}
\ead{fahimthescientist@gmail.com}

\author[1]{Faisal Muhammad Shah} [orcid= 0000-0002-5118-8571]
\ead{faisal.cse@aust.edu}

\cortext[cor1]{Corresponding author}

\begin{abstract}
Early detection of vision-threatening conditions such as diabetic retinopathy, glaucoma, and age-related macular degeneration depends on retinal fundus image analysis, but manual assessment is slow and expert-dependent. Automated convolutional neural networks classify fundus images accurately yet act as black boxes, and existing retinal vessel segmentation methods lose discriminative power under pathology and seldom exploit attention or transformer backbones. Using the public FIVES and DRIVE fundus datasets, we develop a two-pipeline framework that pairs interpretable four-class disease classification with attention- and transformer-based vessel segmentation, organised in three stages: (1) FIVES images are augmented by rotation and horizontal and vertical flips and used to fine-tune eight ImageNet-pretrained CNNs: ResNet101, DenseNet169, Xception, InceptionV3, DenseNet121, InceptionResNetV2, ResNet50, and EfficientNetB0. (2) Five gradient-based explanation methods, Grad-CAM, Grad-CAM++, Score-CAM, Faster Score-CAM, and Layer-CAM, are computed on the final convolutional block of each classifier and compared qualitatively across architectures. (3) Ten U-Net variants are benchmarked for vessel segmentation: TransUNet (hybrid CNN--Transformer encoder) and Attention U-Net (gated skip connections), evaluated with ResNet50V2, ResNet101V2, and ResNet152V2 backbones, along with additional Attention U-Net configurations using DenseNet backbones, and the fully transformer-based Swin-UNet. ResNet101 gives the highest classification accuracy: 94.17\% (F1 0.942) $>$ 88.33\% for EfficientNetB0. For segmentation, the architecture ranking is consistent on both datasets: Attention U-Net $>$ TransUNet $>$ Swin-UNet. The strongest configuration is Attention U-Net with a ResNet101V2 backbone: FIVES IoU 0.722, Dice 0.838; DRIVE IoU 0.648, Dice 0.787, lifting DRIVE IoU 60.80 $\rightarrow$ 64.83 over a prior custom U-Net. Among explanation methods, Score-CAM gives the most spatially focused maps. Across both tasks, deep residual backbones outperform transformer-only designs, and gated skip connections contribute more than transformer encoders to vessel delineation under pathology.
\end{abstract}

\begin{keywords}
Retinal fundus imaging \sep Diabetic retinopathy \sep Explainable AI \sep Grad-CAM \sep Score-CAM \sep Convolutional neural networks \sep Retinal blood vessel segmentation \sep Attention U-Net \sep TransUNet \sep Swin-UNet \sep Medical image analysis
\end{keywords}

\maketitle
\section{Introduction}
The fundus, the posterior segment of the eye, contains the retina, choroid, photoreceptors, retinal vasculature, and optic nerve head, and its appearance carries evidence of ocular and systemic disease. Retinal fundus imaging is a modality for detecting and monitoring vision-threatening disease, making it well suited to routine examination and population screening programmes, and it underpins the diagnosis and follow-up of diabetic retinopathy (DR), glaucoma, and age-related macular degeneration (AMD) \cite{Intro1} \cite{Intro7} \cite{Intro8}. These conditions account for a large share of avoidable blindness, and each is easier to treat when identified early. AMD is a chronic degenerative disease of the macula, marked by drusen deposition in its dry form and abnormal choroidal vessel growth in its wet form, and is a leading cause of irreversible central vision loss in older adults. Glaucoma progresses through optic nerve head damage associated with elevated intraocular pressure and is asymptomatic until an advanced stage, so early detection is essential. DR damages the retinal microvasculature and remains a major cause of preventable blindness, for which timely intervention is critical \cite{Intro3} \cite{Intro9}.

Clinical assessment of these conditions combines fundus photography with ancillary imaging such as optical coherence tomography (OCT) for cross-sectional macular analysis, fluorescein angiography for vascular leakage, and tonometry for intraocular pressure \cite{saba}. Because early disease is frequently silent, screening is needed to identify patients before vision is lost; however, manual grading of fundus images is time-consuming, requires trained specialists, and is subject to inter-observer variability. Deep learning has therefore become central to scalable screening and monitoring \cite{Intro2} \cite{Intro6} \cite{Intro4}. Two problems dominate this setting: classifying the disease present in an image, and segmenting the retinal vasculature, the latter serving as a diagnostic biomarker and as a preprocessing step for downstream analysis \cite{Intro11} \cite{Intro12}.

Both problems expose a common weakness of current models. Convolutional neural networks (CNNs) achieve stro\-ng classification accuracy but act as opaque predictors, limiting clinical trust; explainable AI (XAI) methods that attribute a prediction to image regions are used to address this \cite{Intro5}. For vessel segmentation, encoder-decoder CNNs such as U-Net are effective but lose discriminative power with pathology and thin, low-contrast vessels, motivating architectures that model longer-range context through attention gates or transformer encoders. These two lines of work have largely advanced in isolation, and it is not established whether design choices that help one task also help the other.

Prior studies show both the progress and the remaining gaps. Yeonwoo et al. \cite{Intro1} classified fundus image laterality with a CNN, reaching 99\% training accuracy on 25{,}911 images and using Grad-CAM to align model attention with clinical reasoning, but evaluated no other attribution method. Yong et al. \cite{Intro4} predicted age and sex from retinal features, again relying solely on Grad-CAM. Yan et al. \cite{Intro7} applied deep CNNs to fundus image categorisation at high accuracy but reported no interpretability analysis. For segmentation, Zhiyuan et al. \cite{Intro13} proposed a task-driven generative adversarial network that improved accuracy, sensitivity, specificity, and AUC over classical methods, and Abdulsahib et al. \cite{Intro14} introduced a fully automatic vessel segmentation pipeline with strong results on DRIVE and HRF; neither examined transformer or attention-based backbones. More recent work narrows these gaps individually but not jointly. A lightweight attention-augmented CNN pairs multiclass screening with Grad-CAM and Grad-CAM++ evidence \cite{xai2025light}, and a DenseNet121 glaucoma classifier uses Grad-CAM for lesion-level explanation \cite{xai2025glaucoma}, yet both rely on a single family of gradient attributions and a single backbone. On the segmentation side, Swin-transformer encoder-decoders \cite{seg2025swin} and gated fusion U-Nets \cite{seg2025gated} report competitive DRIVE accuracy but are each evaluated within one architecture family and dataset regime, and interpretable multi-task frameworks that jointly segment lesions and classify disease \cite{xai2025joint} still fix a single segmentation backbone. Across this literature, interpretability is typically limited to one XAI method, segmentation rarely benefits from attention or transformer designs, and no study varies the XAI method, the segmentation paradigm, and the backbone together on more than one dataset.

We address these gaps by studying the two problems in parallel. For classification, we train eight pre-trained CNNs to categorise retinal fundus images into four classes and attribute each prediction with five gradient-based XAI methods, so that classification accuracy and explanation behaviour are examined on the models. For segmentation, we evaluate ten U-Net variants that span attention gates (Attention U-Net), a hybrid CNN-transformer encoder (TransUNet), and a transformer design (Swin-Unet), each paired with several residual or dense backbones. Classification is scored with six metrics covering overall accuracy, the precision-recall balance, and probability calibration, and segmentation with five metrics covering region overlap, boundary agreement, and pixel-level accuracy; the segmentation models are evaluated on two datasets that differ in size and resolution so that the conclusions are not specific to one dataset. The two pipelines are developed separately, but because they use the same backbone families we also compare their outcomes to see whether the architectural choices that help classification also help segmentation.

Our study is organised around four research questions, each targeting one aspect of the two pipelines.
\begin{itemize}
    \item \textbf{RQ1.} How do widely used pre-trained CNN backbones compare for retinal fundus disease classification, and which architectural families are most effective?
    \item \textbf{RQ2.} How do gradient-based Explainable AI methods differ in the quality and localisation of the visual evidence they produce, and how does this depend on the classifier used?
    \item \textbf{RQ3.} How do attention-based and transformer-based U-Net architectures compare with conventional encoder-decoder designs for retinal blood vessel segmentation, and what role does the backbone play?
    \item \textbf{RQ4.} Do the architectural choices that improve classification also improve segmentation, that is, do the findings transfer across the two tasks?
\end{itemize}

\noindent Our contributions are summarised as follows:
\begin{itemize}
    \item We compare eight pre-trained CNN models, namely ResNet101, DenseNet169, Xception, InceptionV3, Den\-seNet121, InceptionResNetV2, ResNet50, and EfficientNetB0, for four-class retinal fundus classification on the FIVES dataset.
    \item We apply five gradient-based XAI methods, specifically Grad-CAM, Grad-CAM++, Score-CAM, Faster Score-CAM, and Layer CAM, to every trained classifier and compare the spatial fidelity of their saliency maps across architectures.
    \item We evaluate ten U-Net variants for retinal blood vessel segmentation on the DRIVE and FIVES datasets: TransUNet and Attention U-Net with DenseNet201, DenseNet169, DenseNet121, ResNet50V2, ResNet10\-1V2, and ResNet152V2 backbones, together with Swin-Unet.
    \item We assess each task with several complementary metrics, six for classification and five for segmentation, covering overall accuracy, probability calibration, region overlap, and boundary agreement.
    \item We compare the classification and segmentation results to identify which backbone and interpretability choices transfer across the two tasks, and we position the strongest segmentation configuration against prior work on DRIVE and FIVES.
\end{itemize}

The remainder of the paper is organised as follows. Section \ref{Rel} reviews related work; Section \ref{datas} describes the datasets; Section \ref{BS} presents background on gradient-based Explainable AI, pre-trained CNNs, and segmentation models; Section \ref{EVM} defines the evaluation metrics for classification and segmentation; Section \ref{methodology} details the proposed approach; Section \ref{exp} reports the experimental setup and hyper-parameter settings; Section \ref{resu} interprets the results; Section \ref{findings} discusses the Explainable AI comparison, the cross-study performance analysis, and the cross-task findings; Section \ref{limitation} and Section \ref{future} outline the limitations and directions for future work; and Section \ref{conclusion} concludes.

\section{Related Works}\label{Rel}
Prior work relevant to this study falls into three directions: deep learning for fundus disease classification, explainable AI for fundus image classifiers, and retinal blood vessel segmentation. We discuss each in turn, group similar methods rather than describing papers individually, and close each direction with the limitation that motivates our study.

\subsection{Retinal Fundus Classification without Explainable AI}
Automated fundus disease classification has largely converged on transfer learning with deep convolutional networks, and existing studies differ mainly in how the pre-trained backbone is used. One group treats the backbone as a fixed feature extractor and couples it with a separate classifier: Muhammad et al. \cite{Related1} combine VGG-19 features with PCA and SVD selection on a 35{,}126-image Kaggle set, and Rohit et al. \cite{Related6} feed CNN descriptors of glaucoma images from ORIGA and DRISHTI-GS to a bank of classical classifiers, of which logistic regression is the strongest. A second group fine-tunes the backbone end to end: Al-Omaisi et al. \cite{Relatedwork4} compare ResNet-101, ResNet-50, and VGG-16 for diabetic retinopathy on a 1{,}607-image hospital set, and Yuhang et al. \cite{Related5} classify tessellated, normal, and degenerative fundi with Inception V3 and ResNet-50. Across both groups the reported accuracies are high, typically above 90\%, and deeper residual or inception backbones are consistently preferred over shallower ones.

These results are difficult to compare because the studies differ in nearly every uncontrolled variable. Datasets range from a few hundred to tens of thousands of images, class definitions vary from binary DR grading to multi-disease labels, and evaluation splits and metrics are rarely standardised, so a high accuracy on one private set cannot be ordered against a high accuracy on another. The feature-extractor pipelines introduce a second design choice, the downstream classifier, whose contribution is seldom separated from that of the backbone. More importantly, none of these studies reports which image regions drive a prediction, so a clinically plausible accuracy is no evidence that the model attends to disease-relevant structure rather than to acquisition artefacts. This lack of transparency is the limitation that explainable classification, discussed next, attempts to address, and it is a central concern of the classification pipeline in this paper.

\subsection{Retinal Fundus Classification with Explainable AI}
Explainable retinal fundus classification has increasingly incorporated post-hoc attribution methods to reveal the visual evidence underlying model predictions, with Grad-CAM and its variants being among the most widely used approaches. Some studies use attribution primarily to validate an existing classifier: Yeonwoo et al. \cite{Intro1} show with activation maps that a laterality classifier attends to the optic disc and vascular arcades in a manner consistent with clinical interpretation, while Kyoung et al. \cite{Intro6} apply the same approach across ResNet-50, VGG-19, and Inception v3 for nine-way disease prediction. Other studies focus primarily on classifier performance, with interpretability serving a secondary role: Yusaku et al. \cite{RelatedWork2} contrast a CNN with an SVM for cross-population DR grading, and Gongpeng et al. \cite{RelatedWork3} compare EfficientNet-B7, DenseNet, and ResNet-101 on ultra-widefield images. More recent work integrates interpretability more directly into the prediction pipeline, coupling ten-class screening with per-prediction Grad-CAM and Grad-CAM++ evidence in a lightweight attention-augmented network that achieves 87.9\% accuracy and a macro-F1 of 0.880 \cite{xai2025light}, while other approaches produce lesion segmentations alongside disease labels to enable cross-validation between the predicted disease and localized retinal abnormalities \cite{xai2025joint}.

Despite this progress, the interpretability component of these studies remains narrow. Almost all rely on a single attribution method, usually vanilla Grad-CAM, and treat its output as reference truth rather than comparing it against alternatives such as Score-CAM or Layer-CAM that differ in resolution and in sensitivity to gradient noise. The explanation is also typically demonstrated on one backbone, so it is unknown whether a method that localises well on a strong network degrades on a weaker one, or whether the ranking of explanation methods is stable across architectures. Finally, explanation quality is judged qualitatively, by visual agreement with clinical expectation, without a consistent criterion. The literature therefore offers little guidance on which explanation method to pair with which classifier, which is the question the interpretability analysis in this paper examines by applying five gradient-based methods to each of eight backbones.

\begin{table*}
\caption{Key findings from prior work on retinal fundus classification, with and without Explainable AI.}\label{relatedtable}
\begin{tabular*}{\textwidth}{@{\extracolsep{\fill}}p{2.1cm}p{2.7cm}p{1.1cm}p{2.5cm}p{3.2cm}@{}}
\toprule
\textbf{Types} & \textbf{Authors} & \textbf{Year} & \textbf{Models Employed} & \textbf{Performance}\\
\midrule
\textbf{Classification without XAI} & Muhammad et al. \cite{Related1} & 2018 & VGG19 & Accuracy 92.21\%\\
\addlinespace
 & Al-Omaisi et al. \cite{Relatedwork4} & 2022 & ResNet-101, ResNet-50, VggNet-16 & Accuracy: ResNet-101: 98.82\%, ResNet-50: 91.5\% and VggNet-16: 64.11\%\\
\addlinespace
 & Yuhang et al. \cite{Related5} & 2023 & InceptionV3 and ResNet-50 & Accuracy: ResNet-50: 93.81\%, InceptionV3: 91.76\%\\
\addlinespace
 & Rohit et al. \cite{Related6} & 2023 & Proposed model & Accuracy: DRISTHI-GS 99\%, ORIGA 99.4\%\\
\midrule
\textbf{Classification with XAI} & Yeonwoo et al. \cite{Intro1} & 2018 & Proposed CNN model & Accuracy: 98.9\%\\
\addlinespace
 & Yusaku et al. \cite{RelatedWork2} & 2020 & SVM and NN classifiers & Sensitivity: 81.5\%, Specificity: 71.9\%\\
\addlinespace
 & Kyoung et al. \cite{Intro6} & 2021 & VGG19, Inceptionv3, ResNet50 & Accuracy: VGG19: 82\%, Inceptionv3: 83.40\%, ResNet50: 87.40\%\\
\addlinespace
 & Gongpeng et al. \cite{RelatedWork3} & 2022 & EfficientNet-B7, DenseNet, and ResNet-101 & Average AUC: EfficientNet-B7: 0.9880, ResNet-101: 0.9784, and DenseNet: 0.9830\\
\addlinespace
 & Arnob et al. \cite{xai2025light} & 2025 & Attention-augmented CNN with Grad-CAM/Grad-CAM++ & Accuracy: 87.9\%, Macro-precision: 0.882, Macro-recall: 0.879, Macro-F1: 0.880\\
\bottomrule
\end{tabular*}
\end{table*}

\subsection{Retinal Blood Vessel Segmentation}
Vessel segmentation has moved from hand-designed filters to encoder-decoder networks and, more recently, to attention and transformer models. Classical approaches use directional filtering and morphology, as in the training-free B-COSFIRE pipeline of Wenjing et al. \cite{RF1}. The current mainstream is the U-Net family, extended along several axes: recurrent skip connections in the RU-Net and R2U-Net models of Yun et al. \cite{RF2}, guided filtering to recover down-sampling losses in Pengshuai et al. \cite{RF3}, multi-scale inputs and dense blocks in S. Alex et al. \cite{RF5}, spatial attention with structured dropout in the SA-UNet of Changlu et al. \cite{RF6}, nested skips in the U-Net++ pipeline of Manizheh et al. \cite{RF7}, and residual blocks with full-scale skips in Ko-Wei et al. \cite{RF8}. Generative training has also been explored, with Zhiyuan et al. \cite{Intro13} using a task-driven GAN with multi-scale discriminators. Transformer-based designs are the newest addition, including a hybrid transformer and convolutional dual-path decoding U-Net that reports accuracies of 97.28\%, 97.55\%, and 97.60\% on DRIVE, STARE, and CHASE\_DB1 \cite{seg2024tcddu}, and a hybrid architecture that embeds transformer blocks within a U-Net and reaches a DRIVE AUC of 0.9861 at 96.97\% accuracy \cite{seg2024panet}.

Two limitations run through this literature. First, evaluation is concentrated on a small set of similar benchmarks, chiefly DRIVE, STARE, and CHASE\_DB1, which contain mostly healthy or mildly diseased retinas; performance on images with extensive pathology, where vessels are obscured by exudates or haemorrhage, is rarely measured. Second, each study introduces one architectural idea and compares it against generic U-Net baselines, so the relative value of the main design axes, attention gating versus a transformer encoder versus a purely convolutional decoder, cannot be read from the results, and the interaction of these choices with the CNN backbone is left unexplored. Establishing that comparison on data that also includes diseased retinas is the aim of the segmentation pipeline in this paper.

\begin{table*}[width=\textwidth,cols=5,pos=htbp]
\caption{Retinal blood vessel segmentation methods and their reported results on standard fundus datasets.}\label{relatedtable2}
\begin{tabular*}{\textwidth}{@{\extracolsep{\fill}}p{1.9cm}p{2.6cm}p{1.1cm}p{3.1cm}p{3.1cm}@{}}
\toprule
\textbf{Types} & \textbf{Authors} & \textbf{Year} & \textbf{Models Employed} & \textbf{Performance}\\
\midrule
\textbf{Segmentation} & Ko-Wei et al. \cite{RF8} & 2023 & U-net, U-Net 3+, ResUNet, ResUNet++ & DRIVE dataset IOU: 60.8, Rose dataset IOU: 59.3\\
\addlinespace
 & Wenjing et al. \cite{RF1} & 2022 & B-Cosfire filter, Morphologic filter & DRIVE Sensitivity: 0.7339, Specificity: 0.9847, Accuracy: 0.9604\\
\addlinespace
 & Manizheh et al. \cite{RF7} & 2022 & U-Net++ & Accuracy: 0.989, Sensitivity: 0.941, Specificity: 0.988\\
\addlinespace
 & S. Alex et al. \cite{RF5} & 2022 & U-Net, Vanilla U-Net & Vanilla U-net: ACC: 0.9557, Sen: 0.712, Spec: 0.992\\
\addlinespace
 & Yun et al. \cite{RF2} & 2020 & U-Net, RU-Net, R2U-Net & Optic Cup Segmentation: Accuracy: 0.9950, Dice coefficient: 0.8921\\
\addlinespace
 & Pengshuai et al. \cite{RF3} & 2020 & U-net, DU-net, LaderNet, CE-Net & Accuracy: 0.9709, Sensitivity: 0.8510, Specificity: 0.9916\\
\addlinespace
 & Zhiyuan et al. \cite{Intro13} & 2020 & U-Net & Accuracy: 0.9683, Sensitivity: 0.8066, Specificity: 0.9897\\
\addlinespace
 & Changlu et al. \cite{RF6} & 2020 & U-net, U-net+SA, SD-Unet, SA-Unet & Accuracy: 0.9698, AUC: 0.9864, F1: 0.8263\\
\addlinespace
 & TCDDU-Net \cite{seg2024tcddu} & 2024 & Transformer + convolutional dual-path decoding U-Net & Accuracy: DRIVE 0.9728, STARE 0.9755, CHASE\_DB1 0.9760\\
\addlinespace
 & PA-Net \cite{seg2024panet} & 2024 & Hybrid CNN-transformer U-Net & DRIVE AUC: 0.9861, Sensitivity: 0.8489, Accuracy: 0.9697\\
\bottomrule
\end{tabular*}
\end{table*}

\subsection{Position of the Proposed Work}
The prior work reviewed above shares a common shortcoming: each study varies a single factor, an architecture, an explanation method, or a backbone, and evaluates it under conditions that are not shared with neighbouring studies. Classification work rarely reports where models look, interpretability work rarely varies the explanation method or the backbone, and segmentation work rarely tests attention and transformer designs against each other or on pathological images. This paper addresses that gap by holding the data and evaluation fixed while varying these factors in a controlled way: eight backbones for four-class classification, five gradient-based explanation methods applied to each of them, and ten U-Net variants spanning attention, hybrid, and transformer designs for vessel segmentation on two datasets that include diseased retinas.

\section{Dataset Description}\label{datas}
\begin{figure*}[pos=htbp]
  \centering
  \begin{subfigure}[b]{0.45\textwidth}
    \includegraphics[width=\textwidth]{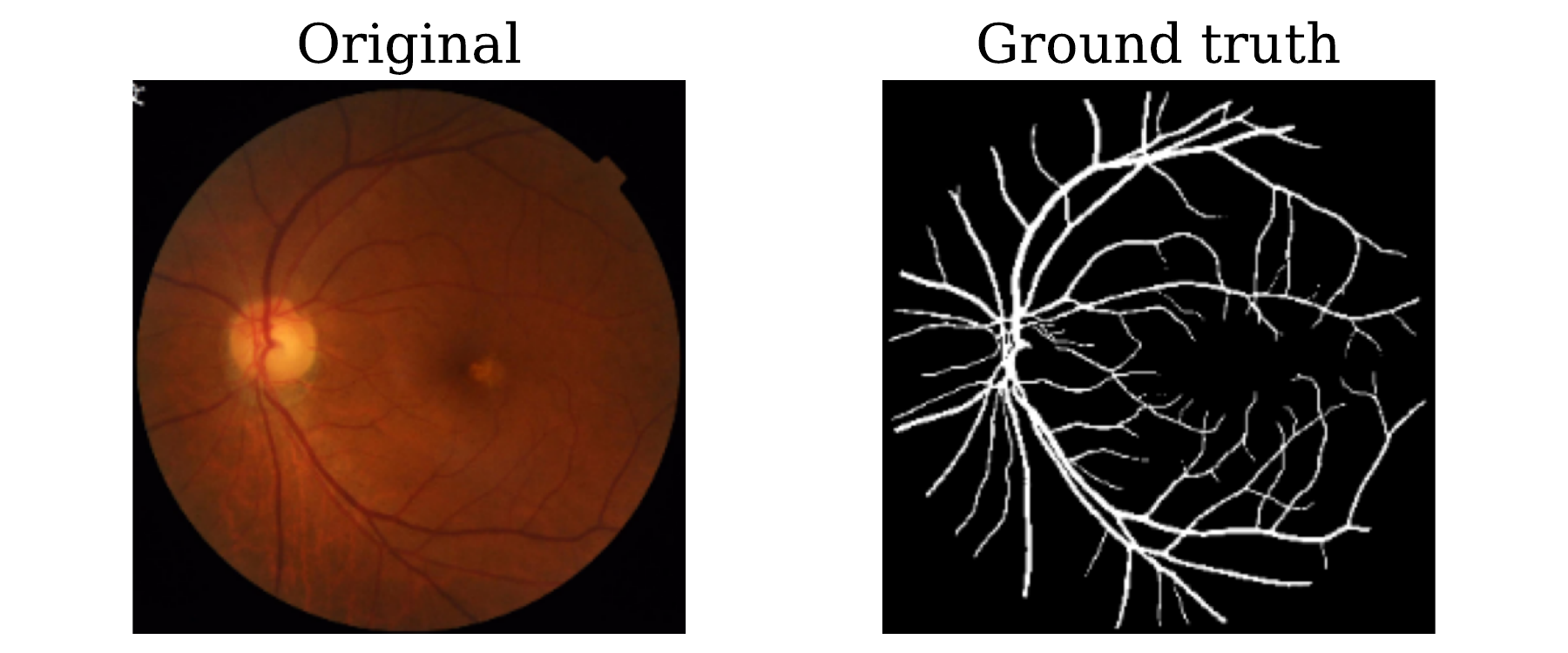}
    \caption{AMD}
    \label{fig:fives_amd}
  \end{subfigure}
  \hfill
  \begin{subfigure}[b]{0.45\textwidth}
    \includegraphics[width=\textwidth]{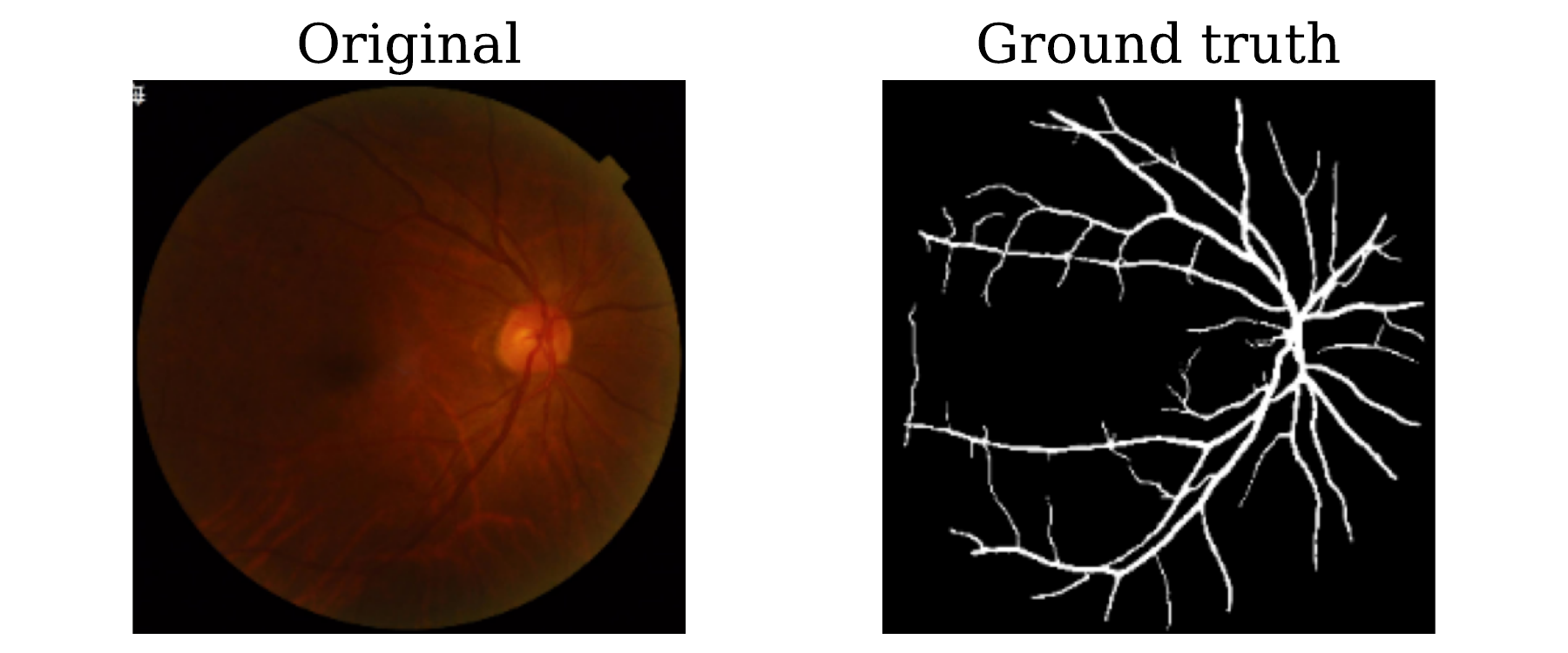}
    \caption{DR}
    \label{fig:fives_dr}
  \end{subfigure}

  \medskip

  \begin{subfigure}[b]{0.45\textwidth}
    \includegraphics[width=\textwidth]{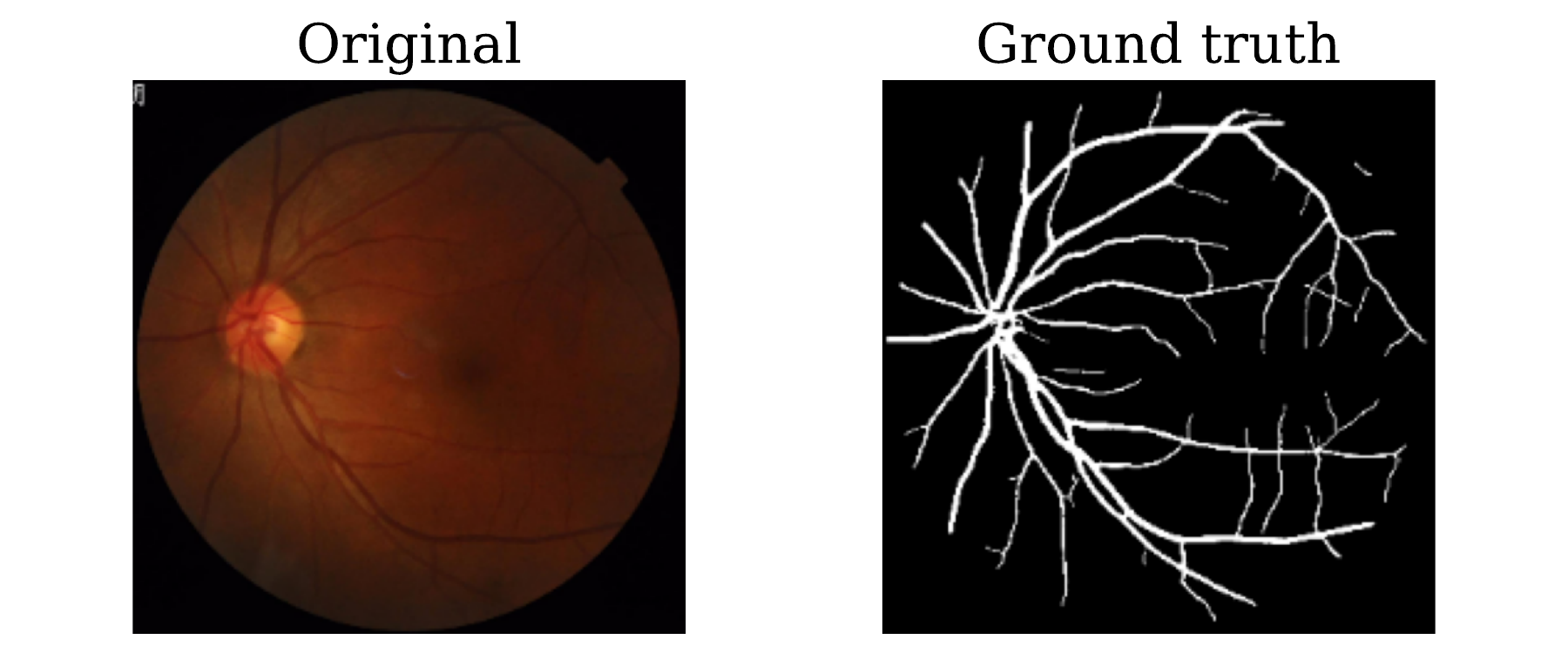}
    \caption{Glaucoma}
    \label{fig:fives_glaucoma}
  \end{subfigure}
  \hfill
  \begin{subfigure}[b]{0.45\textwidth}
    \includegraphics[width=\textwidth]{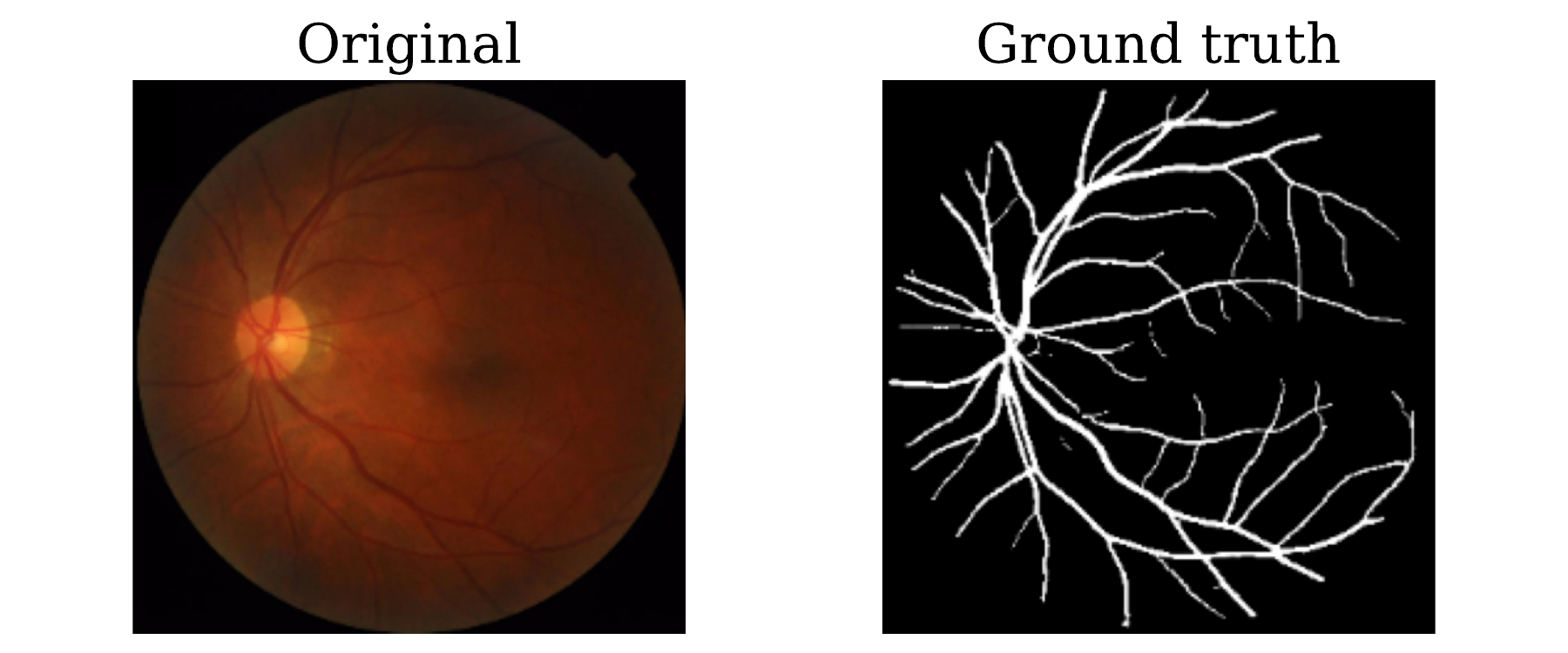}
    \caption{Normal}
    \label{fig:fives_normal}
  \end{subfigure}
  \caption{Representative fundus images from the four categories of the FIVES dataset.}
  \label{fig:Fives_images}
\end{figure*}

\begin{figure*}[pos=htbp]
  \centering
  \begin{subfigure}[b]{0.45\textwidth}
    \includegraphics[width=\textwidth]{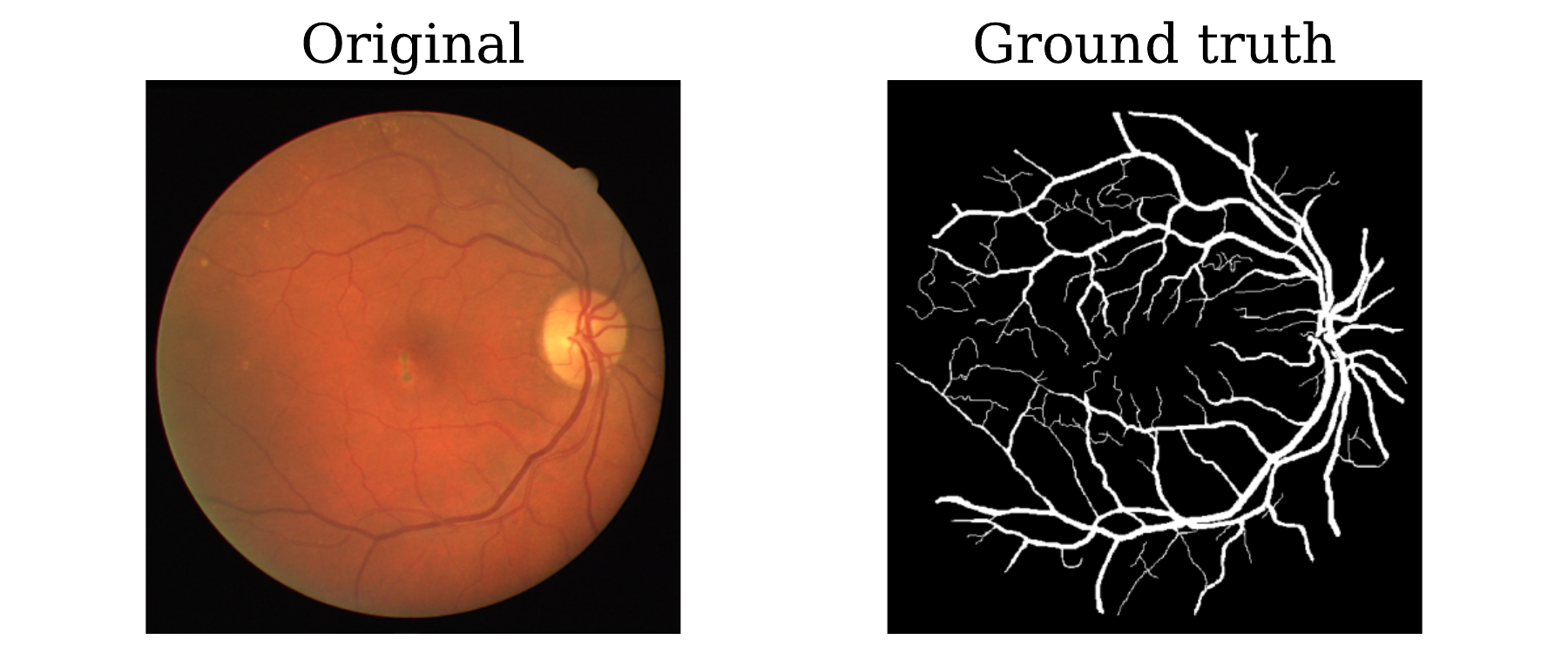}
    \caption{Diseased}
    \label{fig:drive_dis1}
  \end{subfigure}
  \hfill
  \begin{subfigure}[b]{0.45\textwidth}
    \includegraphics[width=\textwidth]{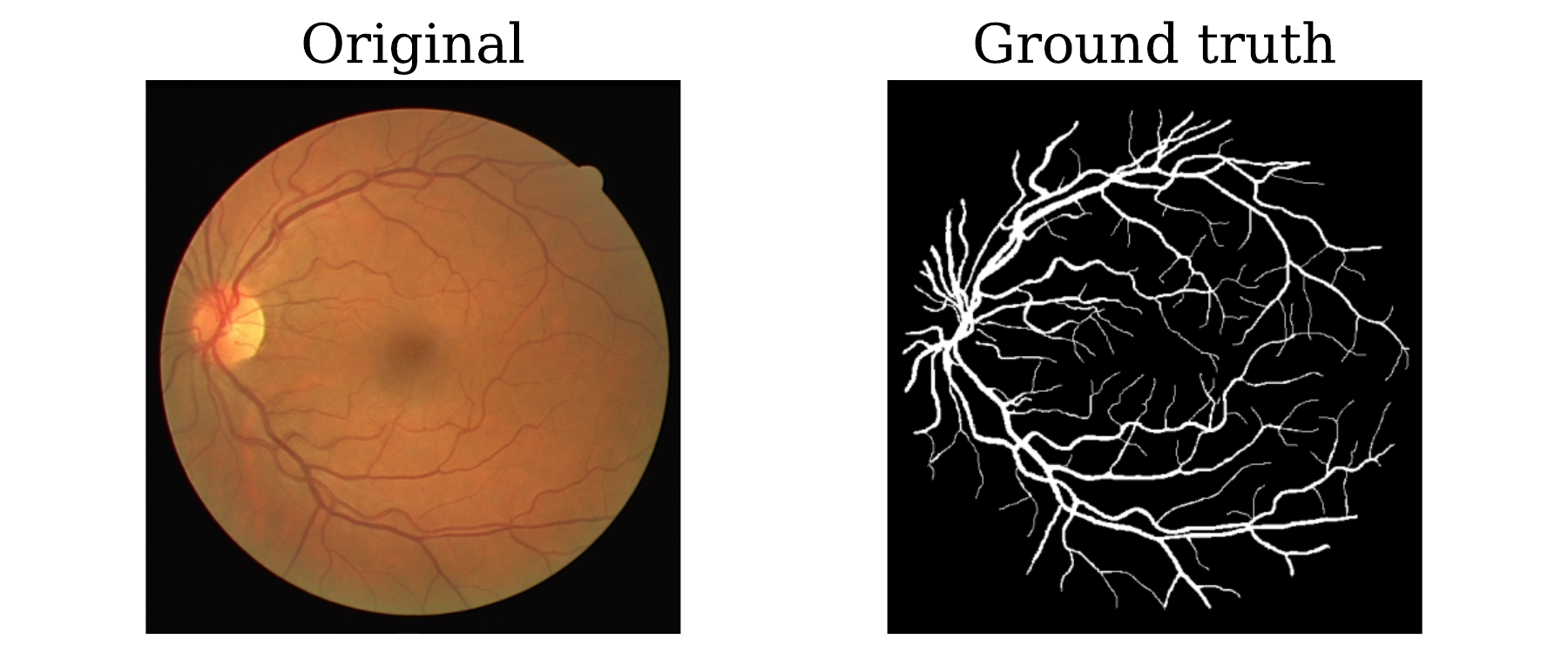}
    \caption{Healthy}
    \label{fig:drive_healthy1}
  \end{subfigure}

  \medskip

  \begin{subfigure}[b]{0.45\textwidth}
    \includegraphics[width=\textwidth]{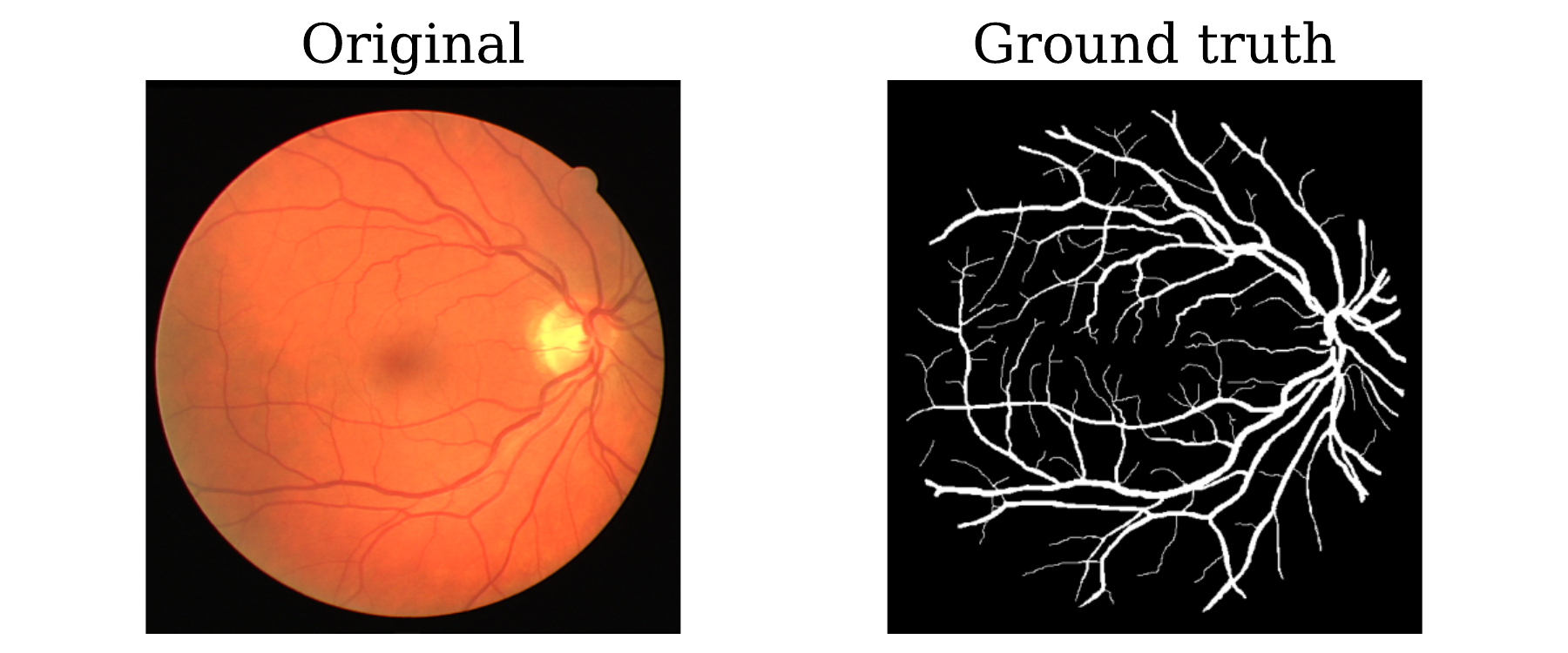}
    \caption{Healthy}
    \label{fig:drive_healthy2}
  \end{subfigure}
  \hfill
  \begin{subfigure}[b]{0.45\textwidth}
    \includegraphics[width=\textwidth]{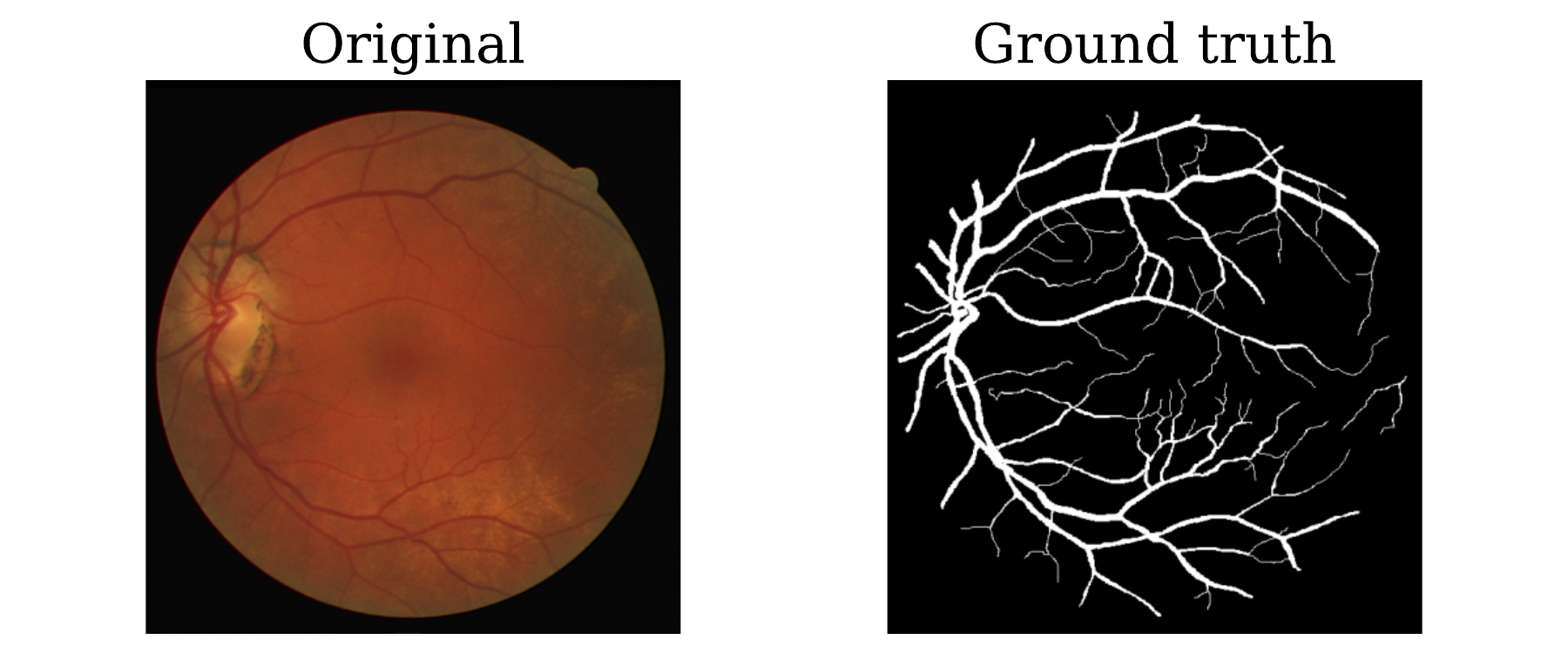}
    \caption{Diseased}
    \label{fig:drive_dis2}
  \end{subfigure}
  \caption{Representative fundus images from the DRIVE dataset.}
  \label{fig:Drive_images}
\end{figure*}

This study relies on two publicly available retinal fundus datasets that were created and annotated by other research groups, FIVES \cite{Fives} and DRIVE \cite{Drive}. We did not collect any images or produce any annotations; both datasets are used as released, with their original images and their authors' expert labels. FIVES supports the classification pipeline, because it is the only large fundus collection that provides a disease label, a pixel-level vessel mask, and an image-quality score for every image. Both FIVES and DRIVE support the segmentation pipeline: DRIVE is the standard benchmark for retinal vessel extraction and allows comparison with prior work, while FIVES adds higher-resolution images and a substantial number of pathological cases. Table \ref{dataset} summarises the two datasets, and representative images are shown in Figure \ref{fig:Fives_images} and Figure \ref{fig:Drive_images}.

\subsection{FIVES}
The Fundus Image Vessel Segmentation (FIVES) dataset \cite{Fives} contains 800 colour fundus photographs stored as uncompressed PNG files at a resolution of $2048 \times 2048$ pixels. The images are evenly divided among four categories, 200 images each of normal retinas, age-related macular degeneration (AMD), diabetic retinopathy (DR), and glaucoma, so the class distribution is balanced and 600 of the 800 images show pathology. Every image carries three annotations produced by the FIVES authors: a disease category label, a pixel-wise binary annotation of the retinal vasculature, and an image-quality score. According to the original publication, the vessel masks were drawn by trained medical annotators and were subsequently checked and corrected by experienced ophthalmologists. FIVES is released with a predefined split of 600 training and 200 test images, with the four categories represented equally in both partitions, and we adopt this split without modification.

\subsection{DRIVE}
The Digital Retinal Images for Vessel Extraction (DRIVE) dataset \cite{Drive}, introduced by Staal et al., contains 40 colour fundus photographs collected during a diabetic retinopathy screening programme in the Netherlands. From 400 screened photographs, 40 were selected, of which 33 show no sign of diabetic retinopathy and 7 show signs of mild early diabetic retinopathy. The images were acquired with a Canon CR5 non-mydriatic 3CCD camera at a $45^{\circ}$ field of view, captured at $768 \times 584$ pixels with eight bits per colour channel, and distributed cropped to $565 \times 584$ pixels around the circular field of view. DRIVE provides a fixed split of 20 training and 20 test images. Each image is accompanied by a binary field-of-view mask; the training images have one manual vessel segmentation, and the test images have two independent manual segmentations, the first of which is the standard reference for evaluation.

\subsection{Data Preparation and Usage}
For the classification pipeline, the FIVES disease labels define a four-class problem, and the images are resized to a fixed input size and augmented during training with rotation and horizontal and vertical flipping; the FIVES training and test partitions are used as released, with the 200-image test partition kept fixed. Since FIVES does not provide a dedicated validation partition, a stratified 10\% subset of the 600-image training partition (60 images) is held out for validation, leaving 540 images for training. This yields an overall split of 540 training (67.5\%), 60 validation (7.5\%), and 200 test (25\%) images.
For the segmentation pipeline, the FIVES and DRIVE vessel masks are the targets, the DRIVE training and test partitions are used as released, and the FIVES partitions are used for its segmentation task in the same way. All images and masks are resized to a common resolution before training. The training, validation, and test partitions are disjoint at the image level in both pipelines, so no image contributes to both model fitting and evaluation.

\begin{table*}[width=\textwidth,cols=6,pos=htbp]
\caption{Composition, resolution, expert annotations, and role in this study of the FIVES and DRIVE retinal fundus datasets.}\label{dataset}
\begin{tabular*}{\textwidth}{@{\extracolsep{\fill}}p{1.4cm}p{3.4cm}p{1.6cm}p{2.1cm}p{3.0cm}p{2.2cm}@{}}
\toprule
\textbf{Dataset} & \textbf{Composition} & \textbf{Images} & \textbf{Resolution} & \textbf{Annotations} & \textbf{Task in this study}\\
\midrule
FIVES \cite{Fives} & 200 normal, 200 AMD, 200 DR, 200 glaucoma (balanced) & 800 & $2048 \times 2048$ (PNG) & Disease label, pixel-wise vessel mask, image-quality score & Classification and segmentation\\
\addlinespace
DRIVE \cite{Drive} & 33 without DR, 7 with mild early DR & 40 & $565 \times 584$ (captured $768 \times 584$) & Vessel masks, field-of-view masks & Segmentation\\
\bottomrule
\end{tabular*}
\end{table*}

\section{Background Study}\label{BS}
\subsection{Gradient-based Explainable AI}
\subsubsection{Grad-CAM:}
Grad-CAM \cite{Grad-CAM}, or Gradient-weighted Class Activation Mapping, is a technique used in deep learning for visualizing the significant regions of an input image that contribute to a specific class prediction in CNNs. During the forward pass, it computes the gradients of the predicted class score with respect to the feature maps, indicating their sensitivity. These gradients are then weighted based on the global average pooling (GAP) of the gradients for each feature map, generating a weighted sum. The resulting weighted sum is passed through a ReLU activation to create an activation map, highlighting the influential regions of the input image for the predicted class. Grad-CAM provides valuable interpretability by offering a visual representation of the areas crucial for a CNN's decision-making process, aiding in understanding and validating the network's behavior. The formula for Grad-CAM can be stated as follows:

\begin{equation}
\alpha_k^c = \frac{1}{Z}\sum_i\sum_j \frac{\partial Y^c}{\partial A_{ij}^k}
\end{equation}

This equation has the following components:  
\begin{itemize}
    \item $\alpha_k^c$ represents the importance of the feature map $k$ for class $c$.
    \item $Z$ is the normalization factor, often the number of elements in the feature map.
    \item $Y^c$ is the class score before the softmax activation.
    \item $A_{ij}^k$ represents the activation of the $k$th feature map at spatial location $(i,j)$.
\end{itemize}

\subsubsection{Grad-CAM++:}
Grad-CAM++ (Gradient-weighted Class Activation Mapping++) \cite{Gradcampp} extends the Grad-CAM method, providing a means to visualize and comprehend the influential regions within an input image for a CNN prediction. By introducing second-order derivatives, Grad-CAM++ refines interpretability, enhancing accuracy in localizing critical areas. The method begins with a forward pass, where the input image traverses the CNN, yielding final convolutional feature maps that capture channel activations. Gradients of the predicted class score are then computed, elucidating the impact of each feature map on the score. Employing Global Average Pooling on these gradients produces importance weights for each feature map, reducing dimensionality to a single weight per channel. Second-order gradients, calculated through backpropagation, provide additional information about the channels. The ultimate visualization is derived by a weighted sum of feature maps, combining first-order and second-order gradients, delivering a nuanced depiction of the image regions influencing the CNN's decision-making process. The equation for Grad-CAM++ is expressed as follows:

\begin{equation}
\alpha_k^c = \frac{1}{Z}\sum_i\sum_j \frac{\partial^2 Y^c}{\partial A_{ij}^k \partial A_{ij}^k} + \frac{2}{Z}\sum_i\sum_j \frac{\partial Y^c}{\partial A_{ij}^k}
\end{equation}

This equation includes the following elements:  
\begin{itemize}
    \item $\alpha_k^c$ represents the importance of the feature map $k$ for class $c$.
    \item $Z$ is the normalization factor, often the number of elements in the feature map.
    \item $Y^c$ is the class score before the softmax activation.
    \item $A_{ij}^k$ represents the activation of the $k$th feature map at spatial location $(i,j)$.
    \item The first term represents the second-order derivatives, and the second term represents the first-order derivatives.
\end{itemize}

\subsubsection{Score-CAM:}
Score-CAM (Score-weighted Class Activation Mapping) \cite{score} is a technique designed for elucidating the pivotal regions within an input image that significantly influence a neural network's classification output. Tailored for classification tasks, Score-CAM enhances interpretability by incorporating class-specific information. In the method's forward pass, the input image undergoes processing through the neural network, yielding final convolutional feature maps. Class Activation Maps (CAM) are then computed via global average pooling on these feature maps, highlighting spatial locations crucial to the class score. Subsequently, class scores are generated by linearly combining CAM values with weights learned from the fully connected layer. Gradients of the class score with respect to the feature maps are calculated, guiding the determination of weights. The final visualization is derived by a weighted sum of feature maps, utilizing these computed gradients as weights, providing a refined and localized interpretation of the neural network's predictions in the context of classification tasks such as Retina Fundus Image Classification. The mathematical expression for Score-CAM is represented as:
\begin{equation}
\alpha_k^c = \frac{1}{Z}\sum_i\sum_j \frac{\partial Y^c}{\partial A_{ij}^k}
\end{equation}

This equation contains: 
\begin{itemize}
    \item $\alpha_k^c$ represents the importance of the feature map $k$ for class $c$.
    \item $Z$ is the normalization factor, often the number of elements in the feature map.
    \item $Y^c$ is the class score before the softmax activation.
    \item $A_{ij}^k$ represents the activation of the $k$th feature map at spatial location $(i,j)$.
\end{itemize}

\subsubsection{Faster Score-CAM:}
Faster Score-CAM \cite{Fscore} for Retina Fundus Image Classification is an optimization technique aimed at enhancing the efficiency of interpreting CNNs in the context of classifying retinal fundus images. Traditional Score-CAM calculates weights for all channels in a chosen feature map, which can be computationally expensive. In response, Faster Score-CAM focuses on channels with high variance, assuming they contain more discriminative information. The process involves calculating mean and variance for each channel, selecting the top N channels with the highest variance, and then using these channels to compute weights and generate a heatmap. The benefits include faster computation, making it more practical for analyzing large datasets of retinal fundus images, while still providing valuable insights into influential image regions for classification tasks like diabetic retinopathy or glaucoma. However, choosing the optimal value for N involves a trade-off between speed and capturing informative details in the heatmap, requiring careful experimentation. The mathematical formulation for Faster Score-CAM is provided as:

\begin{equation}
\alpha_k = \frac{\text{Var}(A_k)}{\sum_{j=1}^{N} \text{Var}(A_j)}
\end{equation}

Within the following formula: 
\begin{itemize}
    \item $\alpha_k$ represents the weight assigned to channel $k$.
    \item $\text{Var}(A_k)$ is the variance of activations in channel $k$.
    \item $N$ is the total number of channels in the chosen feature map.
\end{itemize}

\subsubsection{Layer CAM:}
Layer-CAM \cite{LayerCam} is an insightful visualization technique that enhances our understanding of a CNN decision-making process in the realm of retinal fundus image classification. Operating on the final convolutional layer just before the classification layer, Layer-CAM identifies and highlights critical regions within an image that contribute significantly to a specific class prediction. What sets Layer-CAM apart is its capacity to generate reliable class activation maps not only from the final convolutional layer but also from shallow layers, allowing for a balance between coarse spatial locations and fine-grained object details. By considering the entire convolutional layer, Layer-CAM calculates weights for each channel based on their impact on the class score, creating a heatmap that visually represents the network's attention during classification. Importantly, Layer-CAM's ability to produce class activation maps from different layers, each offering complementary insights, allows for the generation of more precise and integral class-specific object regions. The ease of application to off-the-shelf CNN-based image classifiers, without altering network architectures or back-propagation methods, adds to Layer-CAM's convenience and broad applicability, making it a valuable tool for various image analysis tasks. The mathematical representation of Layer-CAM is as follows:

\begin{equation}
S_c(x) = \text{ReLU}\left( \sum_{c=1}^{C} \left( W_c \times \text{Score}(c, F(x)) \right) \right)
\end{equation}

This equation contains the following components:  
\begin{itemize}
    \item $F(x)$ - Output of the final convolutional layer for an input image $x$. This is a 3D tensor with dimensions $(H, W, C)$, where $H$ and $W$ are the height and width of the feature map, and $C$ is the number of channels.
    \item $W_c$ - Weight for channel $c$ in the convolutional layer.
    \item $\text{Score}(c, F(x))$ - Function that computes the contribution of channel $c$ to the class score for a particular class. This can be a simple linear function or a more complex function depending on the model architecture.
    \item $S_c(x)$ - Class activation map for class $c$.
\end{itemize}

\subsection{Pre-trained Convolutional Neural Networks}
\subsubsection{ResNet101:}
ResNet101 \cite{Restnet101}, a key member of Residual Networks (ResNets), tackles the challenge of training deep neural networks by introducing residual learning. It uses skip connections to smooth gradient flow during backpropagation. With 101 layers, its architecture comprises residual blocks containing multiple convolutional layers and shortcut connections, which can perform identity mapping or involve a 1x1 convolution for dimension matching. Employing a bottleneck architecture in its residual blocks, ResNet101 combines 1x1, 3x3, and 1x1 convolutions to reduce computational complexity while maintaining expressive power, effectively capturing complex features.
\subsubsection{DenseNet169:}
DenseNet169 \cite{Dense169} is a convolutional neural network belonging to DenseNets, designed to overcome challenges in deep networks. It features dense connectivity, where every layer connects to every other layer, promoting direct information flow and enhancing gradient flow during training. The architecture incorporates bottleneck layers within dense blocks, reducing computational complexity while maintaining feature richness. Dense connectivity fosters efficient feature reuse, aiding in meaningful hierarchical feature extraction. Transition blocks control network growth and spatial dimensions.

\subsubsection{Xception:}
Xception \cite{Xception} is a Google-developed convolutional neural network architecture, notable for its extreme inception-style design. The name ``Xception" merges ``Extreme" and ``Inception," signifying its innovative approach. Noteworthy features include extreme depth-wise separable convolutions, replacing traditional convolutions for heightened computational efficiency. Drawing inspiration from the Inception architecture, Xception employs depth-wise separable convolutions in multiple parallel paths, reducing parameters and computation. A distinctive global depth-wise separable convolution captures global dependencies, efficiently merging spatial and channel-wise information, thereby enhancing the model's representational capabilities. 

\subsubsection{InceptionV3:}
InceptionV3 \cite{InceptionV3}, part of Google's Inception family, is a CNN architecture celebrated for its efficiency in image classification and computer vision tasks. Key features include the incorporation of the inception module, utilizing parallel convolutional operations with diverse filter sizes (1x1, 3x3, 5x5) and max-pooling to capture multi-scale features and hierarchical representations. To enhance computational efficiency, InceptionV3 employs factorization techniques, breaking down larger convolutions while maintaining pattern capture. The introduction of auxiliary classifiers at intermediate layers aids in addressing the vanishing gradient problem and provides regularization, enhancing overall training and generalization.

\subsubsection{DenseNet121:}
DenseNet121 \cite{Dense121} is a convolutional neural network reno\-wned for its dense connectivity pattern, fostering feature reuse and propagation throughout the network. It consists of 121 layers organized into dense blocks, transition layers, and a final global average pooling layer followed by a softmax classifier. Dense blocks contain densely connected convolutional layers, while transition layers control spatial dimensions and feature map numbers between dense blocks, often involving batch normalization, 1x1 convolutions, and average pooling for dimensionality reduction and downsampling.

\subsubsection{InceptionResNetV2:}
InceptionResNetV2 \cite{inceptionresnetv2}, a robust CNN architecture introduced by Google, seamlessly blends features from Inception and ResNet models. Leveraging the inception module akin to InceptionV3, it employs parallel convolutional operations with diverse filter sizes and max-pooling, facilitating the capture of multi-scale features and hierarchical representations. Integrating residual connections inspired by ResNet, InceptionResNetV2 enables direct information flow between layers, addressing the vanishing gradient problem for effective training of deep networks. The inclusion of bottleneck layers, combining 1x1 and 3x3 convolutions, optimizes computational complexity while preserving expressive power. Further, the architecture incorporates reduction blocks and auxiliary classifiers to downsample spatial dimensions and enhance training robustness, respectively. 
\subsubsection{ResNet50:}
ResNet50 \cite{Resnet50}, part of the ResNet family developed by Microsoft Research, stands out for its effectiveness in training deep neural networks by addressing the vanishing gradient problem. It introduces the concept of residual learning, employing shortcut connections to enable a direct information flow between layers. The bottleneck architecture within its residual blocks, incorporating 1x1, 3x3, and 1x1 convolutions, efficiently reduces computational complexity while preserving expressive power. The ``50" signifies the total number of layers, reflecting a balanced architecture that achieves sufficient depth for capturing intricate features without becoming overly computationally demanding. 
\subsubsection{EfficientNetB0:}
EfficientNetB0 \cite{EﬀicientNetB0}, part of the EfficientNet family, is known for its balanced accuracy and computational efficiency achieved through compound scaling. It uniformly scales depth, width, and resolution, ensuring increased capacity without a significant rise in computational demands. By employing depth-wise separable convolutions, it efficiently splits spatial and channel-wise correlations, reducing parameters and computational costs while maintaining expressive power. The integration of bottleneck blocks, combining 1x1 and 3x3 convolutions, optimizes the trade-off between computational efficiency and feature representation.
\subsection{Segmentation Models}
\subsubsection{TransUNet:}
TransUNet \cite{transunet} stands out as a powerful solution for medical image segmentation, seamlessly integrating the advantages of both U-Net and Transformers. The framework introduces self-attention mechanisms to address feature resolution loss in Transformers, employing a hybrid CNN-Trans\-former architecture. By harmonizing detailed spatial information from CNN features with the global context encoded by Transformers, TransUNet tactically upsamples and combines features, ensuring precise localization within a u-shaped design. Empirical evidence underscores TransUNet's superior performance across diverse medical image segmentation tasks, showcasing the efficacy of our Transformer-based approach. This innovative framework highlights the significance of incorporating low-level features for heightened segmentation accuracy, offering a promising avenue for advanced medical imaging applications.

\subsubsection{Attention U-Net:}
The Attention U-Net \cite{attUNet} represents an advancement in medical image segmentation over the standard U-Net model. Through the incorporation of attention gates (AG), this architecture excels by autonomously prioritizing target structures of diverse shapes and sizes. The AGs effectively eliminate the reliance on explicit external localization modules, streamlining the network. This enhancement, seamlessly integrated into the U-Net framework with minimal computational overhead, heightens model sensitivity and prediction accuracy. Notably, the attention mechanism surpasses previous approaches by introducing grid-based gating, offering improved specificity to local regions. The utilization of a soft-attention technique in a feed-forward CNN for medical imaging is a pioneering feature, replacing hard-attention methods in image classification and external organ localization models.

\subsubsection{Swin-UNET:}
Swin-Unet \cite{swinunet} is a specialized deep learning model tailored for medical image segmentation, diverging from conventional U-Net architectures by integrating the Swin Transformer. Inspired by U-Net, it retains the U-shaped encoder-decoder design with skip connections for holistic information capture. The Swin Transformer in the encoder offers efficiency in processing high-resolution images through parallelized patch-based handling, while excelling at capturing long-range dependencies crucial for understanding intricate medical structures. The decoder, also employing a Swin Transformer, incorporates a ``patch expanding layer" for resolution enhancement during upsampling, aiding in the recovery of spatial details lost in the encoding phase. The use of skip connections ensures direct injection of high-resolution feature maps into the decoder, preserving vital spatial information. Swin-Unet emerges as a potent alternative for medical image segmentation, leveraging the advantages of Swin Transformers for efficient and contextually rich analysis of complex medical data.

\section{Evaluation metrics for both Classification and Segmentation}\label{EVM}
\subsection{Classification Metrics}
\subsubsection{Accuracy:} Accuracy \cite{accu} is a vital metric in Fundus Image Classification as it encapsulates the overall correctness of the model in accurately categorizing fundus images. It serves as a comprehensive gauge of the model's effectiveness in handling various cases. The score can be calculated as:

\begin{equation}
Accuracy = \frac{Number\ of\ Correctly\ Classified\ Fundus\ Images}{Total\ Number\ of\ Fundus\ Images}
\end{equation}

\subsubsection{Precision:} Precision \cite{accu}  is crucial in Fundus Image Classification, emphasizing the model's accuracy in identifying positive cases, such as detecting eye diseases in fundus images. It aims to minimize false positives, enhancing the model's reliability by avoiding unnecessary alarms for healthy conditions. Precision is calculated as follows:

\begin{equation}
Precision = \frac{Correctly\ Classified\ Fundus\ Images}{Total\ Classified\ Fundus\ Images}
\end{equation}

\subsubsection{Recall:} Recall \cite{accu}, also known as Sensitivity or True Positive Rate, is vital in Fundus Image Classification, assessing the model's capability to capture all actual positive instances. It prevents overlooking potential cases of eye diseases in fundus images, emphasizing comprehensive detection. The formula for recall is as follows:

\begin{equation}
Recall = \frac{Correctly\ Classified\ Fundus\ Images}{Total\ Actual\ Fundus\ Images}
\end{equation}

\subsubsection{F1 Score:} The F1 Score \cite{accu} is a crucial metric in Fundus Image Classification, effectively balancing precision and recall. By offering a harmonic mean, it provides a comprehensive evaluation of the model's performance in detecting eye diseases in fundus images. This balance ensures accurate identification of positive cases while capturing all instances, contributing to a robust and reliable classification system. The formula for the F1 Score is:

\begin{equation}
F1 \ Score = \frac{2 \cdot {P} \cdot R}{P + R}
\end{equation}

Where:
\begin{itemize}
    \item $P$ represents the Precision for  Fundus  Images 
    \item $R$ represents the Recall for  Fundus  Images.
\end{itemize}

\subsubsection{Jaccard Score:} Jaccard Score \cite{Jaccard}, also known as Intersection over Union, is important in Fundus Image Classification for measuring the similarity between predicted and actual positive instances within images. Jaccard Score is calculated using the following formula:

\begin{equation}
Jaccard \ Score = \frac{UAP}{IAP}
\end{equation}

Where:
\begin{itemize}
    \item $UAP$ represents Union  Area  of  Predicted  Images  and  Actual   Images 
    \item $IAP$ represents Intersection  Area  of  Predicted  Images and  Actual  Images
\end{itemize}


\subsubsection{Log Loss:} The computation of Log Loss \cite{accu}, also known as Logarithmic Loss, is essential to Fundus Image Classification since it evaluates the model's prediction confidence by taking into account probability values for different classes. For Log Loss, the following formula applies:

\begin{equation}
Log \ Loss = -\frac{1}{N} \sum_{i=1}^{N} \sum_{j=1}^{C} y_{ij} \log(p_{ij})
\end{equation}

where:
\begin{itemize}
    \item $N$ is the total number of fundus images,
    \item $C$ is the total number of classes (e.g., normal, diabetic retinopathy, glaucoma, etc.),
    \item $y_{ij}$ is a binary indicator of whether fundus image $i$ belongs to class $j$,
    \item $p_{ij}$ is the predicted probability that fundus image $i$ belongs to class $j$.
\end{itemize}

\subsection{Segmentation Metrics}
\subsubsection{Intersection over Union (IoU):}
IoU \cite{IOU} measures the overlap between the predicted and ground truth segmentation masks. Specifically, in retinal blood vessel segmentation, IoU quantifies the agreement between the segmented vessels and the actual vessel locations. It is a valuable metric for assessing the spatial accuracy of the segmentation model. The Intersection over Union (IoU) is defined as:

\begin{equation}
IoU = \frac{{Area \ of \ Intersection}}{Area \ of \ Union} 
\end{equation}

\subsubsection{Dice Coefficient:}
Similar to IoU, the Dice Coefficient \cite{dice} evaluates the overlap between predicted and ground truth segmentations. It is particularly useful in the context of retinal blood vessel segmentation as it provides a balanced measure of segmentation accuracy, emphasizing both false positives and false negatives. It is calculated as twice the intersection divided by the sum of areas of the predicted and ground truth masks. The equation for the dice coefficient is as follows:

\begin{equation}
Dice = \frac{{2 \times |A \cap B|}}{{|A| + |B|}}
\end{equation}

Where:
\begin{itemize}
    \item $A$ represents the predicted segmentation mask,
    \item $B$ represents the ground truth segmentation mask.
\end{itemize}

\subsubsection{Mean Pixel Accuracy:}
Mean Pixel Accuracy \cite{MPA} assesses the accuracy of individual pixel-wise classifications in the segmented images. In retinal blood vessel segmentation, this metric measures the overall correctness at the pixel level, offering insights into the model's ability to precisely delineate vessel boundaries. The following formula yields the mean pixel accuracy:
\small{
\begin{equation}
Mean \ Pixel \ Accuracy = \frac{{Number \ of \ correctly \ classified \ pixels}}{{Total \ number \ of \ pixels}}
\end{equation}}

\subsubsection{Mean Modified Hausdorff Distance:}
Mean Modified Hausdorff Distance \cite{MHD} quantifies the dissimilarity between the predicted and ground truth segmentation masks, emphasizing the spatial discrepancies. In retinal blood vessel segmentation, this metric helps evaluate how well the model captures the overall structure and spatial distribution of blood vessels. The equation for the Mean Modified Hausdorff Distance is as follows:

\small{
\begin{equation}
MMHD = \frac{1}{N} \sum_{i=1}^{N} \frac{1}{|B|} \sum_{b \in B} \min_{a \in A} d(a, b)
\end{equation}}

where:
\begin{itemize}
    \item $MMHD$ represents Mean Modified Hausdorff Distance.
    \item $N$ is the total number of ground truth segmentation masks,
    \item $A$ represents the predicted segmentation mask,
    \item $B$ represents the ground truth segmentation mask,
    \item $d(a, b)$ denotes the distance between pixel $a$ in the predicted mask and pixel $b$ in the ground truth mask.
\end{itemize}

\subsubsection{Mean Surface Dice Overlap:}
Mean Surface Dice Overlap \cite{MSD} assesses the agreement between the predicted and ground truth surfaces, providing a comprehensive measure of segmentation accuracy. In retinal blood vessel segmentation, it is particularly valuable for evaluating how well the model captures the 3D structure and complexity of blood vessels. Below is the equation for the Mean Surface Dice Overlap:
\begin{equation}
Mean \ Surface \ Dice \ Overlap = \frac{1}{N} \sum_{i=1}^{N} \frac{{2 \times |A_i \cap B_i|}}{{|A_i| + |B_i|}}
\end{equation}

where:
\begin{itemize}
    \item \(N\) is the total number of samples or images,
    \item \(A_i\) represents the predicted surface for the \(i\)-th sample,
    \item \(B_i\) represents the ground truth surface for the \(i\)-th sample.
\end{itemize}

\section{Proposed Methodology}\label{methodology}
\begin{figure*}
    \centering
    \includegraphics[width=1.0\textwidth]{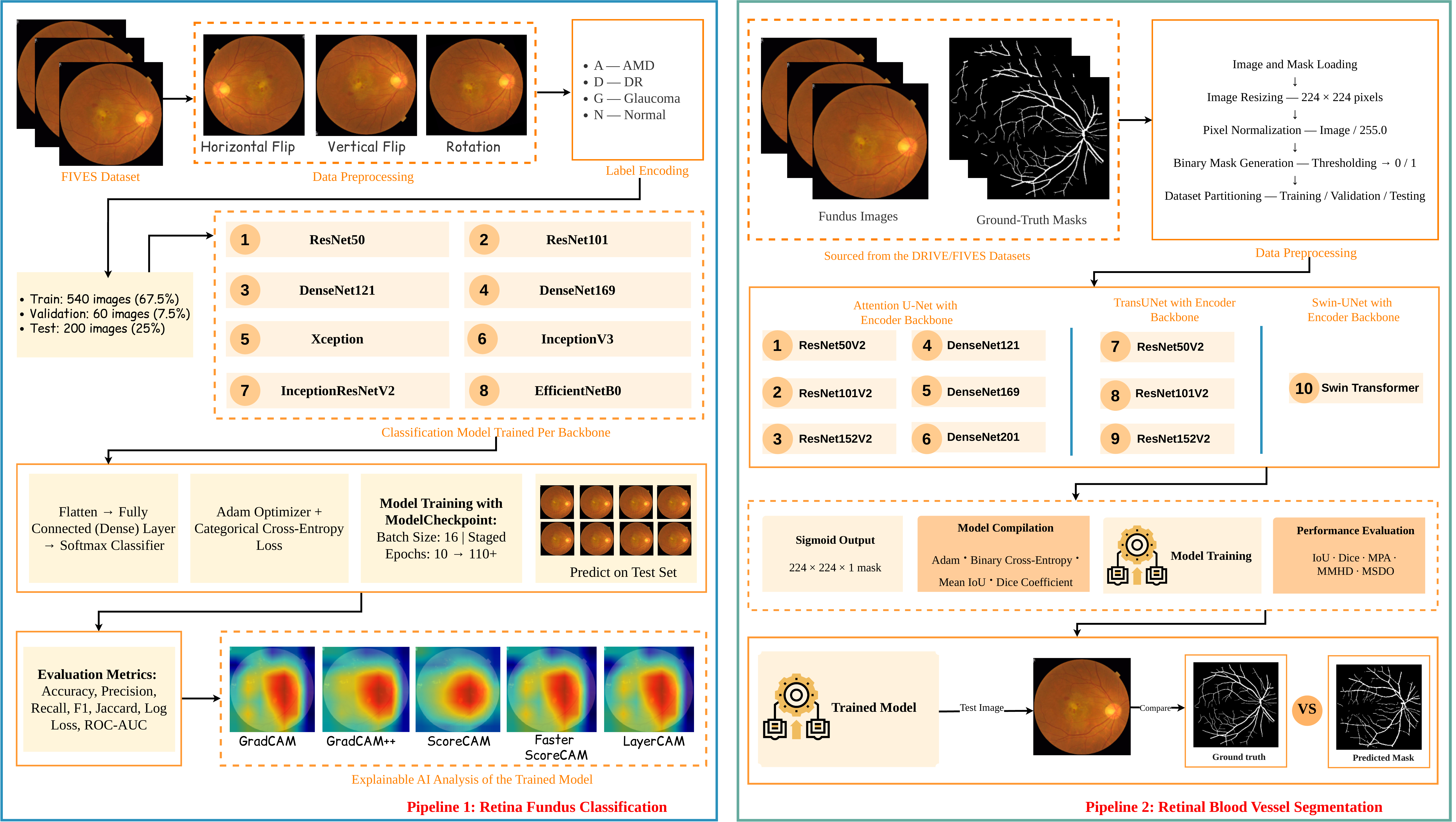}    \caption{The two pipelines for retinal image analysis. Pipeline~1 standardises the fundus images, fine-tunes eight pre-trained backbones for disease classification, and explains each classifier with five gradient-based attribution methods. Pipeline~2 standardises the images and their vessel masks and trains ten U-Net variants for retinal blood vessel segmentation.}
    \label{fig:workflow}
\end{figure*}

This section describes the two pipelines used in this study, from a raw fundus image to the quantities that are scored in the evaluation, and Figure \ref{fig:workflow} gives an overview. \hyperref[pipe1]{Pipeline~1} fine-tunes a set of pre-trained convolutional backbones for disease classification and then explains their decisions with five gradient-based attribution methods. \hyperref[pipe2]{Pipeline~2} trains ten U-Net variants for retinal blood vessel segmentation that differ in their decoder mechanism and encoder backbone. The two pipelines are run independently and share only the convolutional backbone families, which lets us ask whether a backbone that helps one task also helps the other. They meet only in the analysis, where the backbone families used for the two tasks are compared.

\subsection{Pipeline 1: Retinal Fundus Classification}\label{pipe1}
\subsubsection{Data Preprocessing}
Each FIVES image is resized to 299 $\times$ 299 pixels to provide a consistent spatial input to the network. During training, three augmentation operations are applied on the fly: random rotation within $\pm15^\circ$, horizontal flipping, and vertical flipping. Random rotation introduces small variations in image orientation, while horizontal and vertical flipping expose the model to different spatial arrangements of retinal structures. These transformations are applied while preserving the underlying diagnostic information of the fundus images. The training and test partitions described in Section \ref{datas} are used as released, with a stratified validation subset held out from the training partition. The resulting pipeline produces standardized and augmented image-label pairs for the four-class classification task: age-related macular degeneration, diabetic retinopathy, glaucoma, and normal.

\subsubsection{Model Training}
The eight pre-trained backbones introduced in Section \ref{BS} are compared under matched conditions. Training a backbone from scratch is not feasible with a few hundred images per class, which motivates transfer learning: each backbone is initialised with its publicly released pre-trained weights, its original classifier is replaced by a four-way head, and the network is fine-tuned on the augmented training set. A softmax layer turns the outputs into class probabilities, and training minimises the standard multi-class cross-entropy against the ground-truth labels; the log loss reported in Section \ref{EVM} is this same quantity measured on the test set. Optimisation uses the Adam optimiser, with the per-model learning rate, batch size, and epoch count listed in Table \ref{tableparaclass}, and the validation set is used to monitor convergence. Because the optimiser, the augmentation, and the data split are fixed across all eight backbones, with only the per-model schedule allowed to vary, differences in test performance can be attributed mainly to the backbone. This step outputs eight fine-tuned classifiers and their predicted probabilities on the test set.

\subsubsection{Explaining the Classifiers}
The goal of this step is to expose which parts of an input image support a prediction and to compare how five gradient-based attribution methods differ in the evidence they produce, since a high accuracy alone does not show that a model relies on pathology rather than on acquisition artefacts, and prior work has usually fixed on a single method. Given a trained classifier, a test image, and its predicted class, each method reads the activation maps of a chosen convolutional layer and derives an importance weighting over them; the methods differ in how that weighting is obtained: (a) Grad-CAM uses first-order gradients, (b) Grad-CAM++ uses higher-order gradients, (c) Score-CAM uses the class scores obtained when each activation map masks the input, (d) Faster Score-CAM restricts Score-CAM to the highest-variance channels, and (e) Layer CAM uses element-wise positive gradients. The weighting is combined with the activation maps and rectified so that only regions raising the predicted-class score remain, giving a coarse localisation map that is normalised, upsampled to the input resolution, and overlaid on the fundus image, as in Figure \ref{fig:XAI}. Applying every method to every trained backbone, rather than to a single network, is what lets the ranking of attribution methods be tested for dependence on the classifier. This step outputs one overlay per classifier, method, and image.

\subsection{Pipeline 2: Retinal Blood Vessel Segmentation}\label{pipe2}
\subsubsection{Data Preprocessing}
Every FIVES and DRIVE image is resized together with its reference vessel mask to 224 $\times$ 224 pixels. The training and test partitions of Section \ref{datas} are used as released, with a validation subset held out from the training partition where one is available. This step outputs aligned image and mask pairs for both datasets.

\subsubsection{Model Training}
Ten U-Net variants from a public implementation collection are compared. They combine three architecture families, described in Section \ref{BS}, with the encoder backbones listed there: an attention-gated U-Net, a hybrid design that applies transformer layers to convolutional bottleneck features, and a fully transformer U-Net. A plain U-Net decoder tends to lose thin vessels under pathology, which motivates the inclusion of the attention gates and transformer encoders that model longer-range context. Each network produces a per-pixel vessel probability map and is trained by minimising a pixel-wise segmentation loss against the reference mask, using the Adam optimiser with the learning rates, batch sizes, and epoch counts in Table \ref{tableparaFives} for FIVES and Table \ref{tableparaDrive} for DRIVE. At inference the probability map is converted to a binary mask by a fixed threshold. Because all variants share the optimiser and the data split, later differences can be traced to the decoder mechanism and the backbone. This step outputs, for each dataset, ten trained segmentation models and their predicted masks on the test images.

\subsection{Evaluation Protocol}
Both pipelines are evaluated on held-out test data with the metrics defined in Section \ref{EVM}. The classifiers of \hyperref[pipe1]{Pipeline~1} are scored with accuracy, precision, recall, F1 score, Jaccard score, and log loss, which together capture overall correctness, the precision and recall balance, and probability calibration. The segmentation models of \hyperref[pipe2]{Pipeline~2} are scored on the DRIVE and FIVES test sets with intersection over union, the Dice coefficient, mean pixel accuracy, mean modified Hausdorff distance, and mean surface Dice overlap, which together capture region overlap, boundary agreement, and pixel-level correctness. The five attribution methods are compared qualitatively on the two strongest classifiers by visual assessment of how well each heatmap localises disease-relevant structure. No model-specific tuning is applied beyond the hyper-parameter tables, and all models within a task share the same data split, so the comparisons that follow reflect the model rather than the protocol. The experimental environment and the full hyper-parameter settings are reported in Section \ref{exp}.

\section{Experiments}\label{exp}
\subsection{Experimental Setup}
All models were implemented in TensorFlow and trained across three hardware environments. Two were local workstations running Python 3.8.18 with TensorFlow 2.6.0: the first with an NVIDIA GeForce RTX 3050 GPU (compute capability 8.6), an Intel Core i5-9400F CPU, and 16~GB of RAM, and the second with an NVIDIA GeForce RTX 3060~Ti GPU (compute capability 8.6), an Intel Core i5-13600K CPU, and 32~GB of RAM. The third environment was the Kaggle platform, running Python 3.10.13 with TensorFlow 2.15.0 on an NVIDIA Tesla P100 GPU (compute capability 6.0), an Intel Xeon CPU, and 12.72~GB of RAM. The same preprocessing and evaluation code was used in all three environments.

\subsection{Hyper-parameter Settings}
Table \ref{tableparaclass} lists the training settings for the eight pre-trained classifiers of \hyperref[pipe1]{Pipeline~1}. All eight models share a learning rate of 0.001, a batch size of 10, and the Adam optimiser; only the number of training epochs is set per model, ranging from 36 for InceptionResNetV2 to 120 for EfficientNetB0, with the validation loss used to decide when a model had converged. Fixing the learning rate, batch size, and optimiser keeps the training procedure comparable across backbones, while the per-model epoch budget accommodates their different convergence speeds.

\begin{table*}[width=\textwidth,cols=5,pos=htbp]
\caption{Hyperparameter Tuning for Fundus Image Classification Using Pretrained CNNs for FIVES dataset}\label{tableparaclass}
\begin{tabular*}{\textwidth}{@{\extracolsep{\fill}}lllll@{}}
\toprule
\textbf{Models} & \textbf{Learning Rate} & \textbf{Batch Size} & \textbf{Number of Epochs} & \textbf{Optimizer}\\
\midrule
ResNet101 & 0.001 & 10 & 60 & Adam\\
DenseNet169 & 0.001 & 10 & 50 & Adam\\
Xception & 0.001 & 10 & 80 & Adam\\
InceptionV3 & 0.001 & 10 & 70 & Adam\\
DenseNet121 & 0.001 & 10 & 50 & Adam\\
InceptionResNetV2 & 0.001 & 10 & 36 & Adam\\
ResNet50 & 0.001 & 10 & 60 & Adam\\
EfficientNetB0 & 0.001 & 10 & 120 & Adam\\
\bottomrule
\end{tabular*}
\end{table*}

Table \ref{tableparaFives} and Table \ref{tableparaDrive} list the training settings for the ten U-Net variants of \hyperref[pipe2]{Pipeline~2} on the FIVES and DRIVE datasets. All variants use a learning rate of 0.001, a batch size of 4, and the Adam optimiser on both datasets, and again only the epoch count varies. On FIVES the TransUNET variants are trained for 80 epochs, the Attention U-Net variants for 65, and Swin-UNET for 100; on DRIVE the corresponding budgets are 100, 80, and 100 epochs. The smaller batch size relative to the classification pipeline reflects the higher memory cost of the segmentation networks at the input resolution used for this task.

\begin{table*}[width=\textwidth,cols=6,pos=htbp]
\caption{Hyperparameter Tuning for Retinal Blood Vessel Segmentation using U-Net Variants with Different Backbone Architectures for FIVES dataset}\label{tableparaFives}
\begin{tabular*}{\textwidth}{@{\extracolsep{\fill}}llllll@{}}
\toprule
\textbf{Models} & \textbf{Backbone} & \textbf{Learning Rate} & \textbf{Batch Size} & \textbf{Number of Epochs} & \textbf{Optimizer}\\
\midrule
TransUNET & ResNet50V2 & 0.001 & 4 & 80 & Adam\\
 & ResNet101V2 & 0.001 & 4 & 80 & Adam\\
 & ResNet152V2 & 0.001 & 4 & 80 & Adam\\
\midrule
Attention U-Net & ResNet50V2 & 0.001 & 4 & 65 & Adam\\
 & ResNet101V2 & 0.001 & 4 & 65 & Adam\\
 & ResNet152V2 & 0.001 & 4 & 65 & Adam\\
 & DenseNet121 & 0.001 & 4 & 65 & Adam\\
 & DenseNet169 & 0.001 & 4 & 65 & Adam\\
 & DenseNet201 & 0.001 & 4 & 65 & Adam\\
\midrule
Swin-UNET & Swin Transformer & 0.001 & 4 & 100 & Adam\\
\bottomrule
\end{tabular*}
\end{table*}

\begin{table*}[width=\textwidth,cols=6,pos=htbp]
\caption{Hyperparameter Tuning for Retinal Blood Vessel Segmentation using U-Net Variants with Different Backbone Architectures for DRIVE dataset}\label{tableparaDrive}
\begin{tabular*}{\textwidth}{@{\extracolsep{\fill}}llllll@{}}
\toprule
\textbf{Models} & \textbf{Backbone} & \textbf{Learning Rate} & \textbf{Batch Size} & \textbf{Number of Epochs} & \textbf{Optimizer}\\
\midrule
TransUNET & ResNet50V2 & 0.001 & 4 & 100 & Adam\\
 & ResNet101V2 & 0.001 & 4 & 100 & Adam\\
 & ResNet152V2 & 0.001 & 4 & 100 & Adam\\
\midrule
Attention U-Net & ResNet50V2 & 0.001 & 4 & 80 & Adam\\
 & ResNet101V2 & 0.001 & 4 & 80 & Adam\\
 & ResNet152V2 & 0.001 & 4 & 80 & Adam\\
 & DenseNet121 & 0.001 & 4 & 80 & Adam\\
 & DenseNet169 & 0.001 & 4 & 80 & Adam\\
 & DenseNet201 & 0.001 & 4 & 80 & Adam\\
\midrule
Swin-UNET & Swin Transformer & 0.001 & 4 & 100 & Adam\\
\bottomrule
\end{tabular*}
\end{table*}


\section{Results}\label{resu}

\subsection{Fundus Disease Classification}
\begin{figure*}
    \centering
    \begin{subfigure}{0.32\textwidth}
        \includegraphics[width=\linewidth]{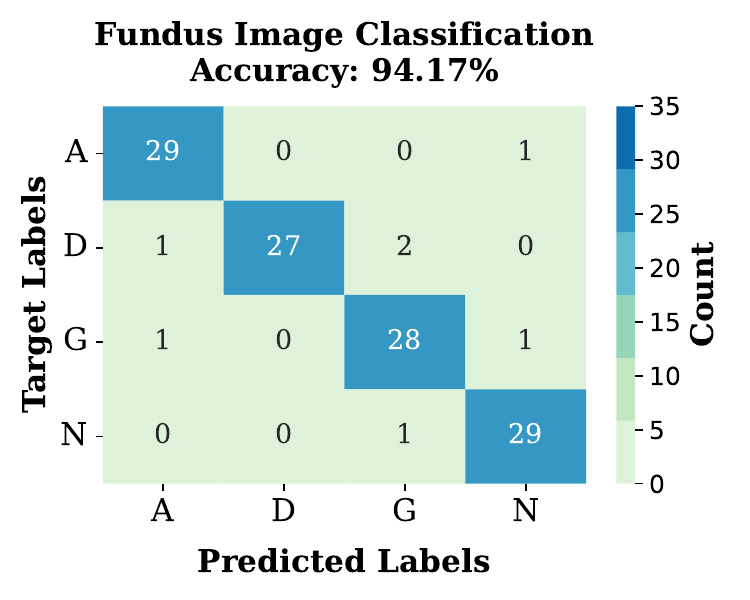}
        \caption{ResNet101}
        \label{fig:subC1}
    \end{subfigure}
    \hfill
    \begin{subfigure}{0.32\textwidth}
        \includegraphics[width=\linewidth]{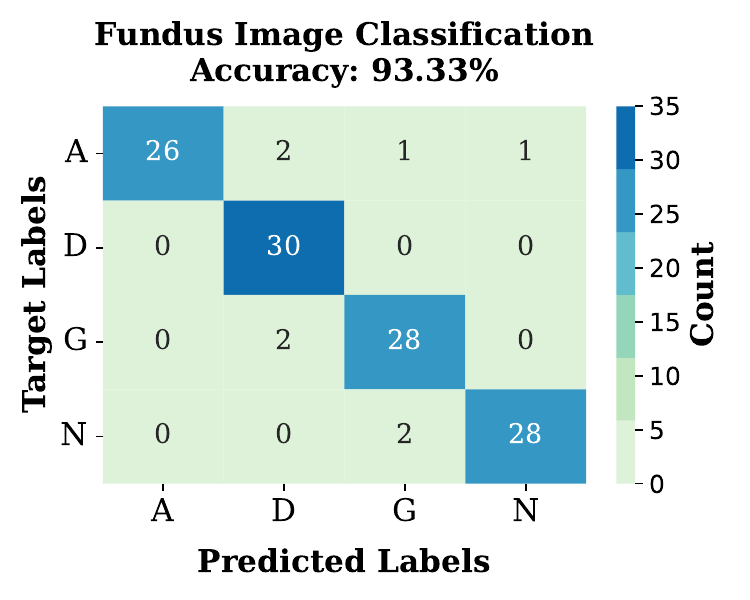}
        \caption{DenseNet169}
        \label{fig:subC2}
    \end{subfigure}
    \hfill
    \begin{subfigure}{0.32\textwidth}
        \includegraphics[width=\linewidth]{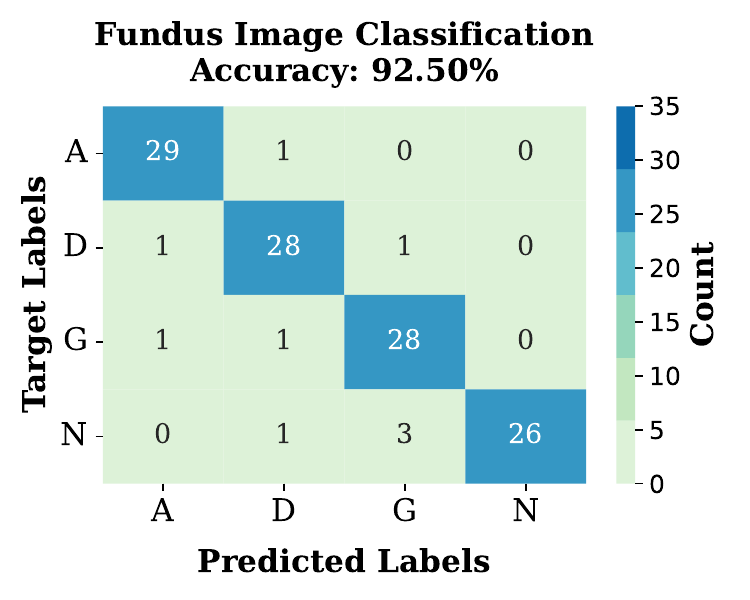}
        \caption{Xception}
        \label{fig:subC3}
    \end{subfigure}
     
     \medskip

    \begin{subfigure}{0.32\textwidth}
        \includegraphics[width=\linewidth]{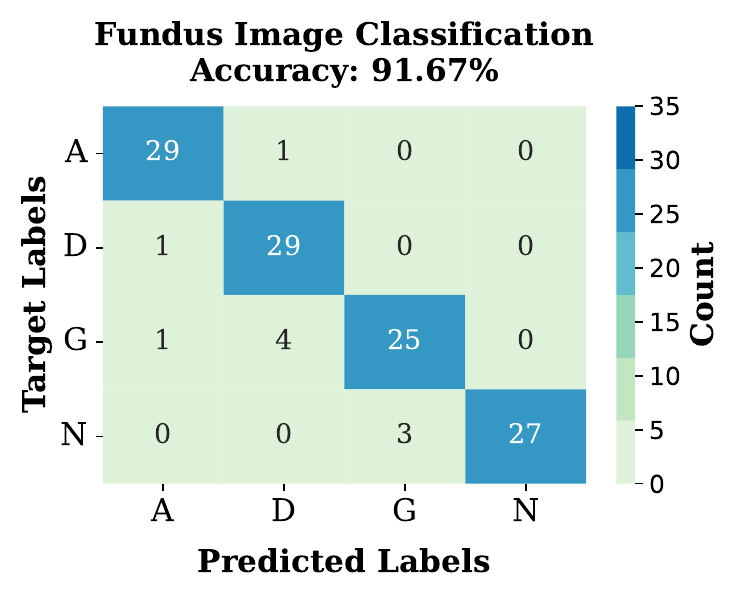}
        \caption{InceptionV3}
        \label{fig:subC4}
    \end{subfigure}
    \hfill
    \begin{subfigure}{0.32\textwidth}
        \includegraphics[width=\linewidth]{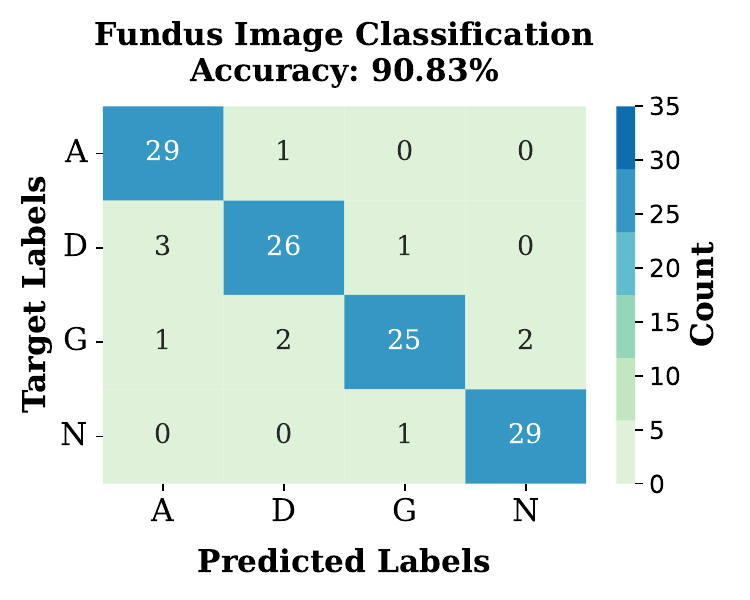}
        \caption{DenseNet121}
        \label{fig:subC5}
    \end{subfigure}
    \hfill
    \begin{subfigure}{0.32\textwidth}
        \includegraphics[width=\linewidth]{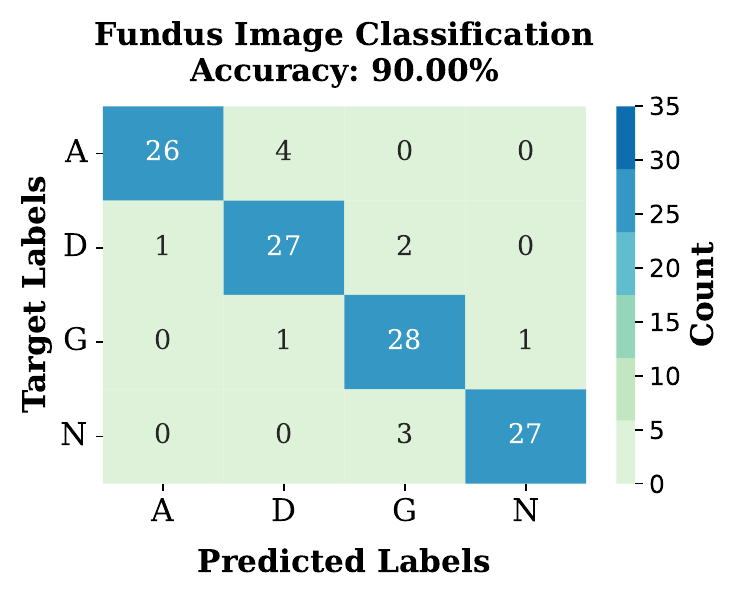}
        \caption{InceptionResNetV2}
        \label{fig:subC6}
    \end{subfigure}
    \medskip

      
    \begin{subfigure}[b]{0.32\textwidth}
        \includegraphics[width=\linewidth]{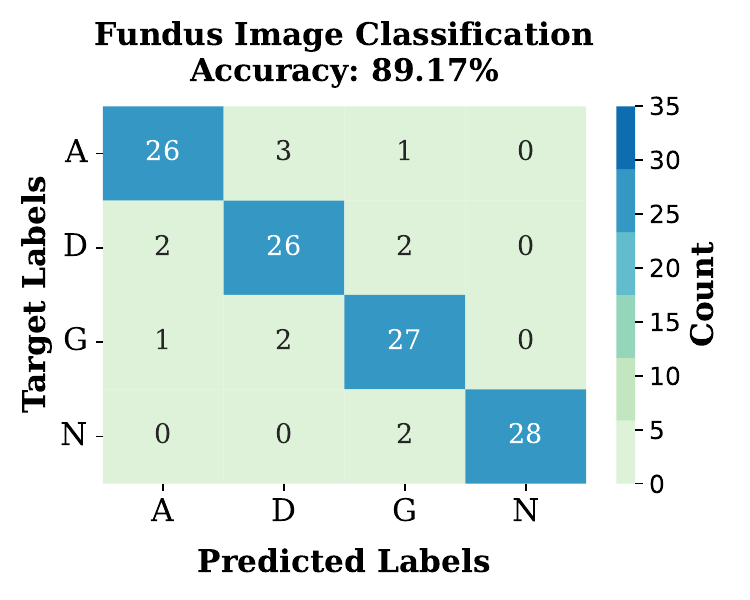}
        \caption{ResNet50}
        \label{fig:subC7}
    \end{subfigure}
     ~
    \begin{subfigure}[b]{0.32\textwidth}
        \includegraphics[width=\linewidth]{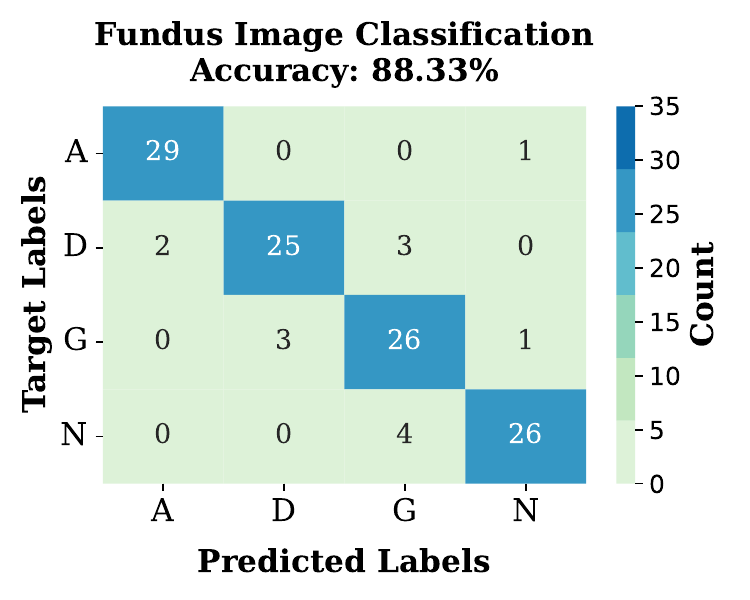}
        \caption{EﬀicientNetB0}
        \label{fig:subC8}
    \end{subfigure}
      

    \caption{Confusion matrices of the eight pre-trained CNN backbones on the FIVES test set.}\label{fig:CM} 
\end{figure*}

\begin{figure*}
    \centering
    \begin{subfigure}{0.32\textwidth}
        \includegraphics[width=\linewidth]{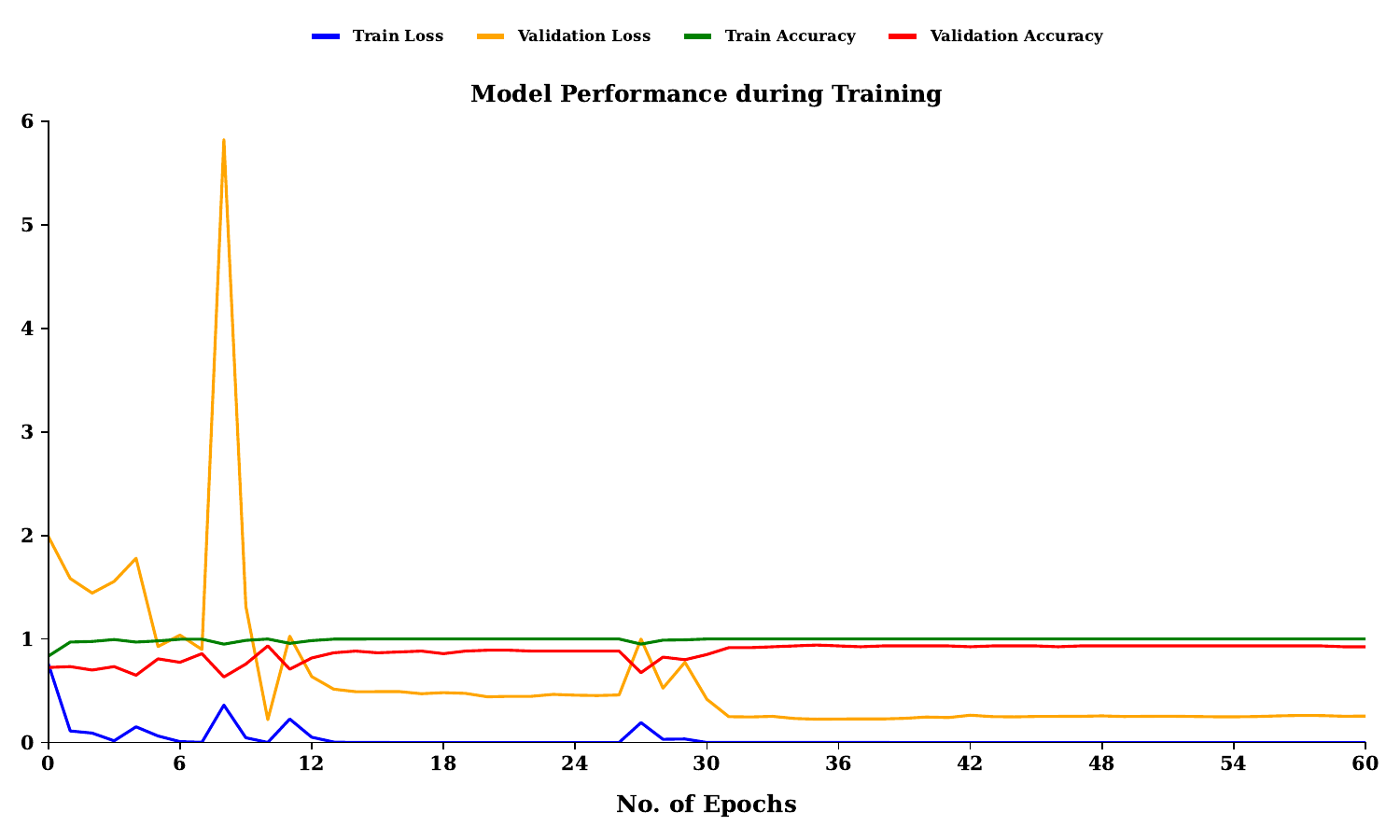}
        \caption{ResNet101}
        \label{fig:subAC1}
    \end{subfigure}
    \hfill
    \begin{subfigure}{0.32\textwidth}
        \includegraphics[width=\linewidth]{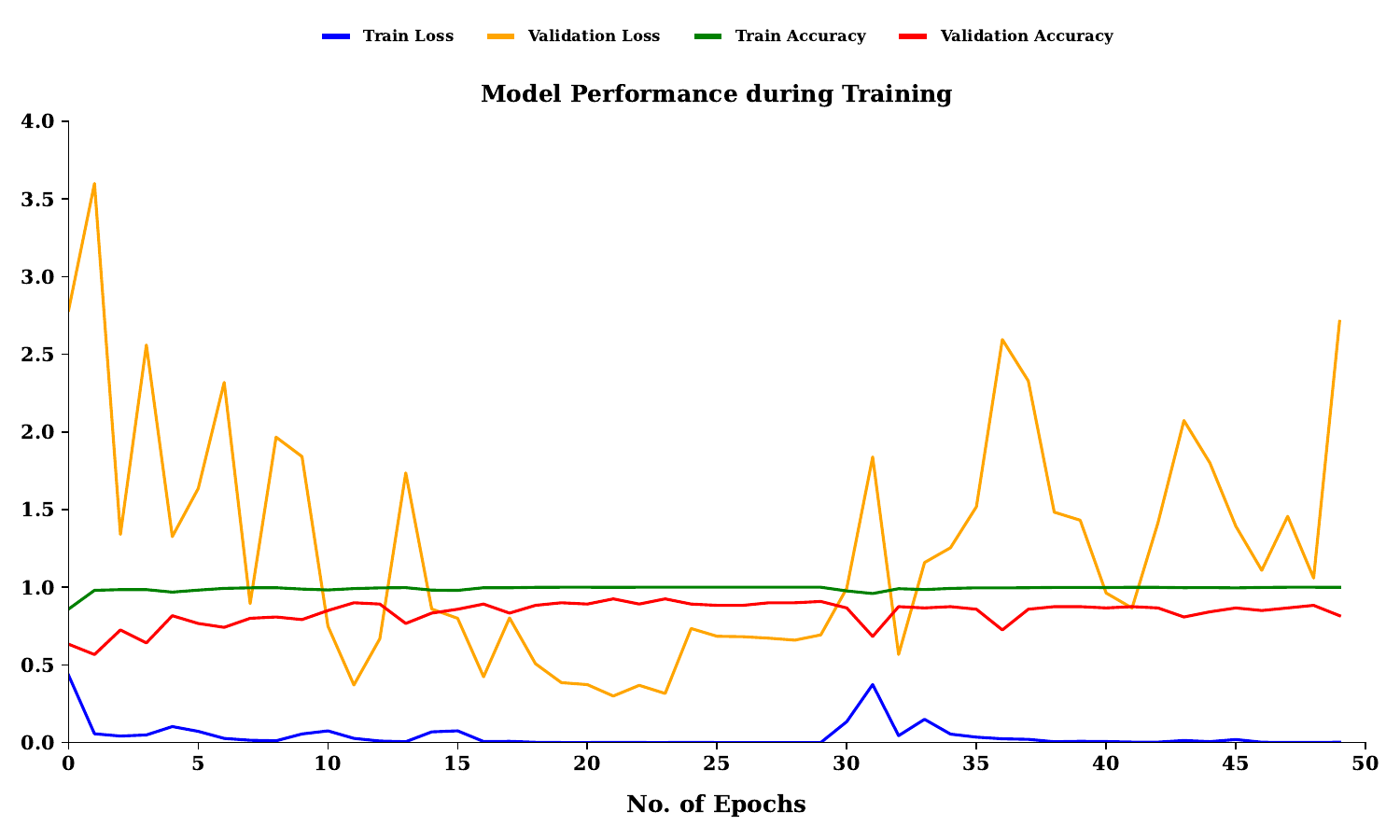}
        \caption{DenseNet169}
        \label{fig:subAC2}
    \end{subfigure}
    \hfill
    \begin{subfigure}{0.32\textwidth}
        \includegraphics[width=\linewidth]{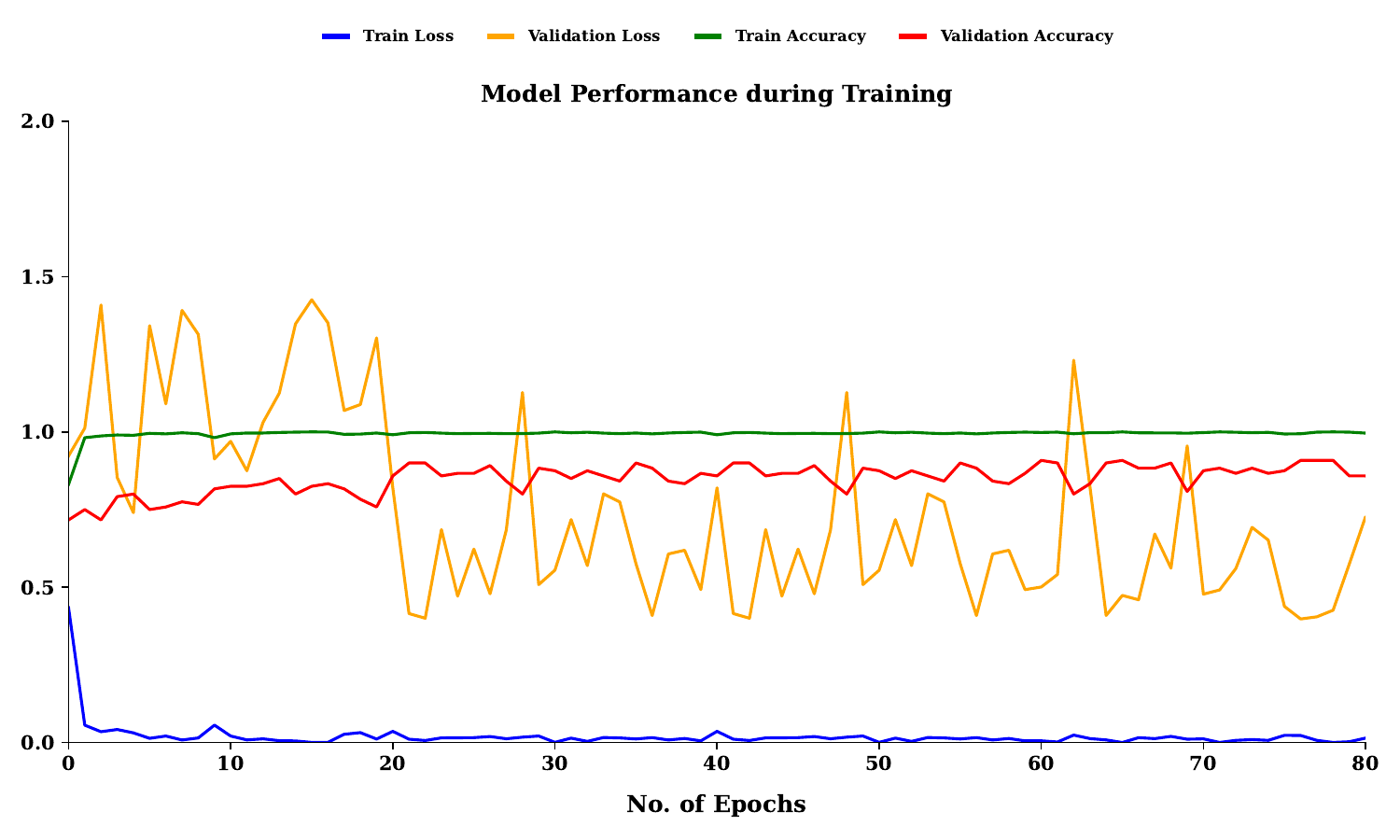}
        \caption{Xception}
        \label{fig:subAC3}
    \end{subfigure}
     \hfill
    \begin{subfigure}{0.32\textwidth}
        \includegraphics[width=\linewidth]{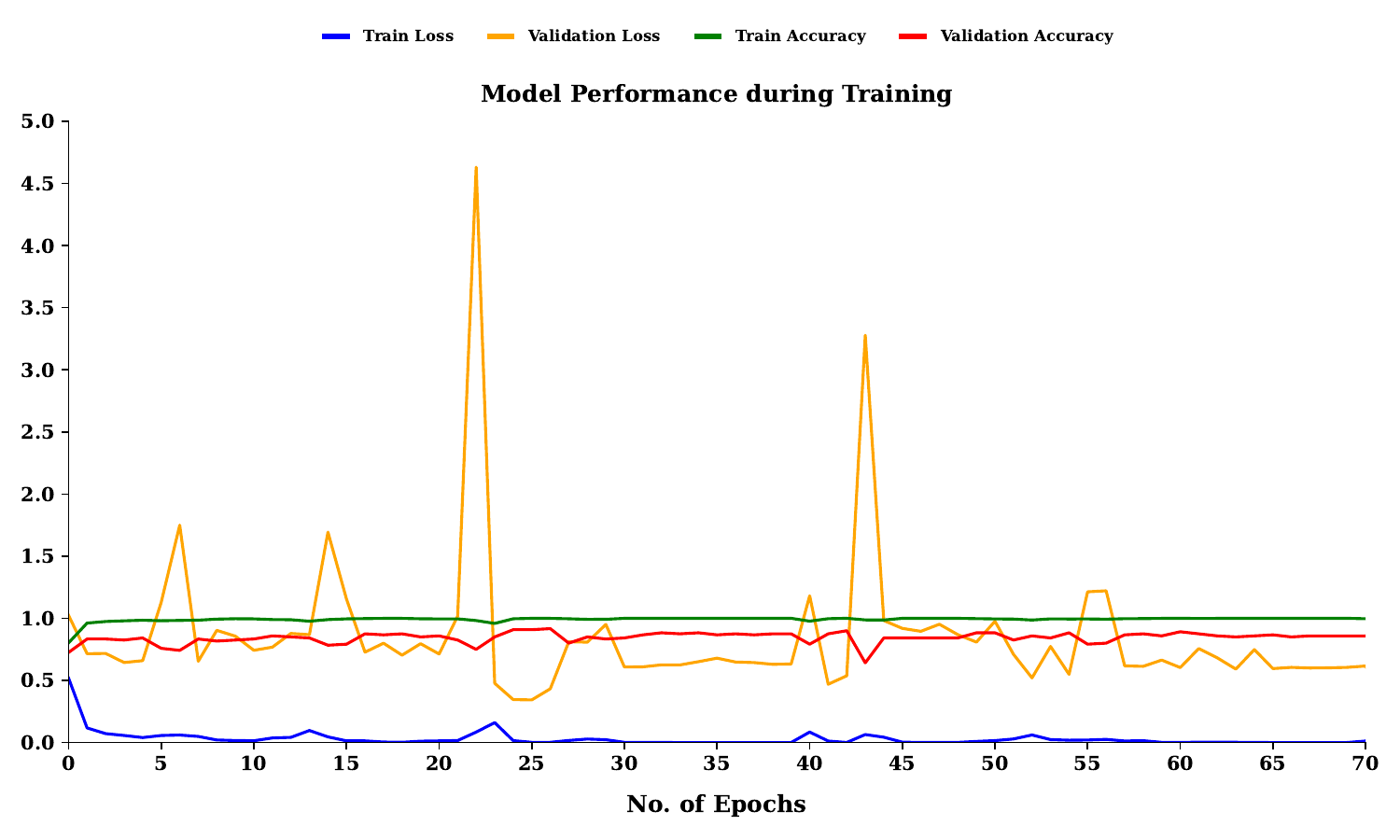}
        \caption{InceptionV3}
        \label{fig:subAC4}
    \end{subfigure}
 \hfill
    \begin{subfigure}{0.32\textwidth}
        \includegraphics[width=\linewidth]{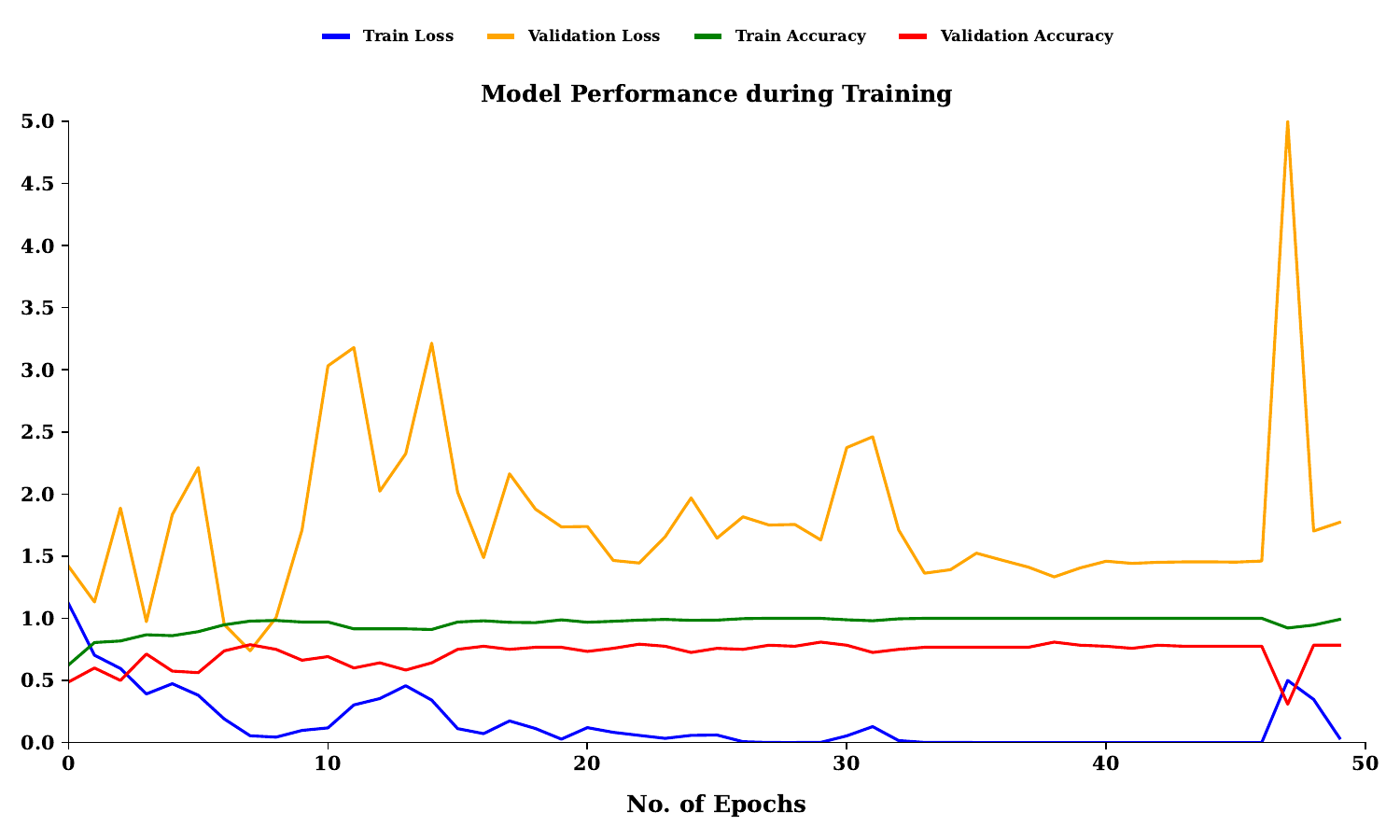}
        \caption{DenseNet121}
        \label{fig:subAC5}
    \end{subfigure}
    \hfill
    \begin{subfigure}{0.32\textwidth}
        \includegraphics[width=\linewidth]{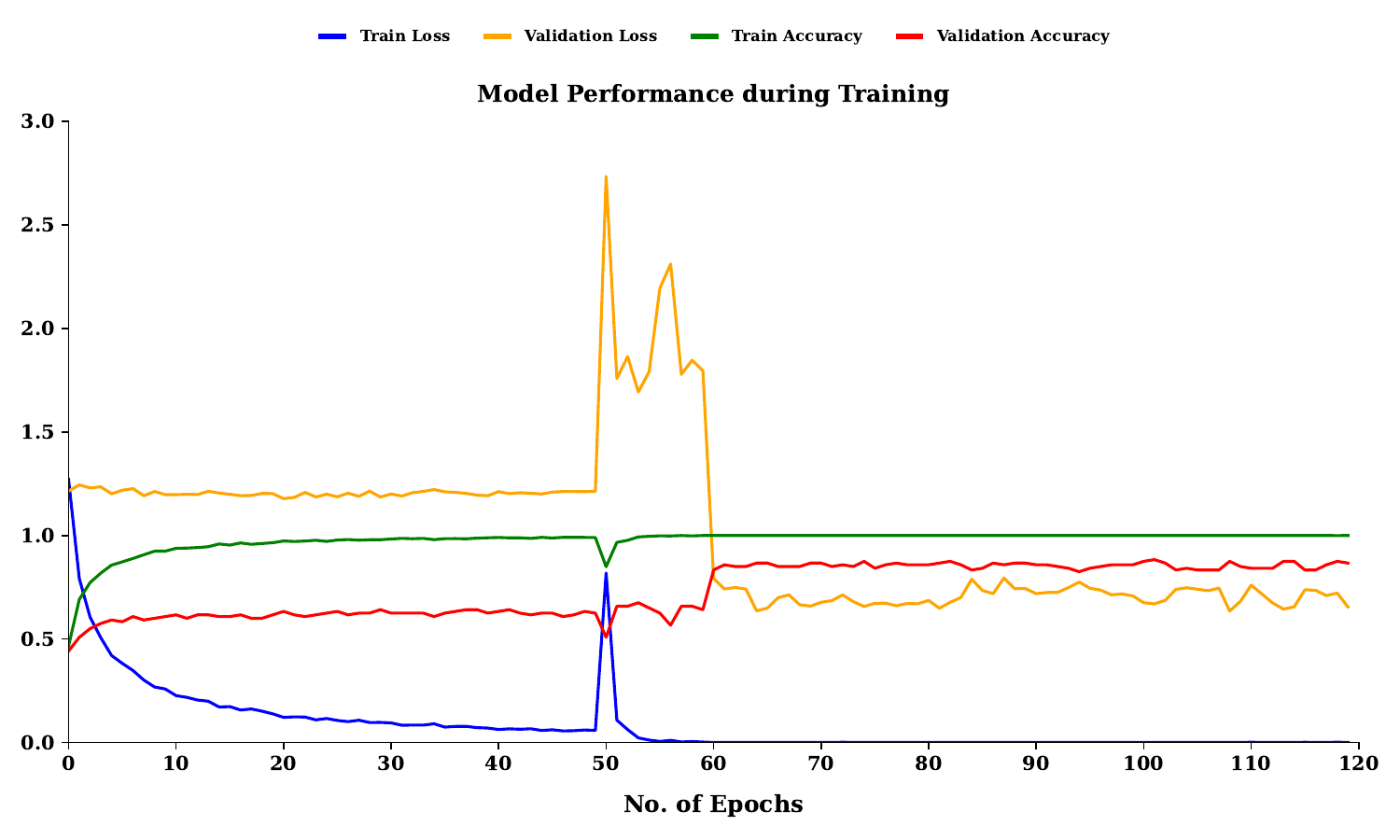}
        \caption{EﬀicientNetB0}
        \label{fig:subAC6}
    \end{subfigure}
    
    \begin{subfigure}{0.32\textwidth}
        \includegraphics[width=\linewidth]{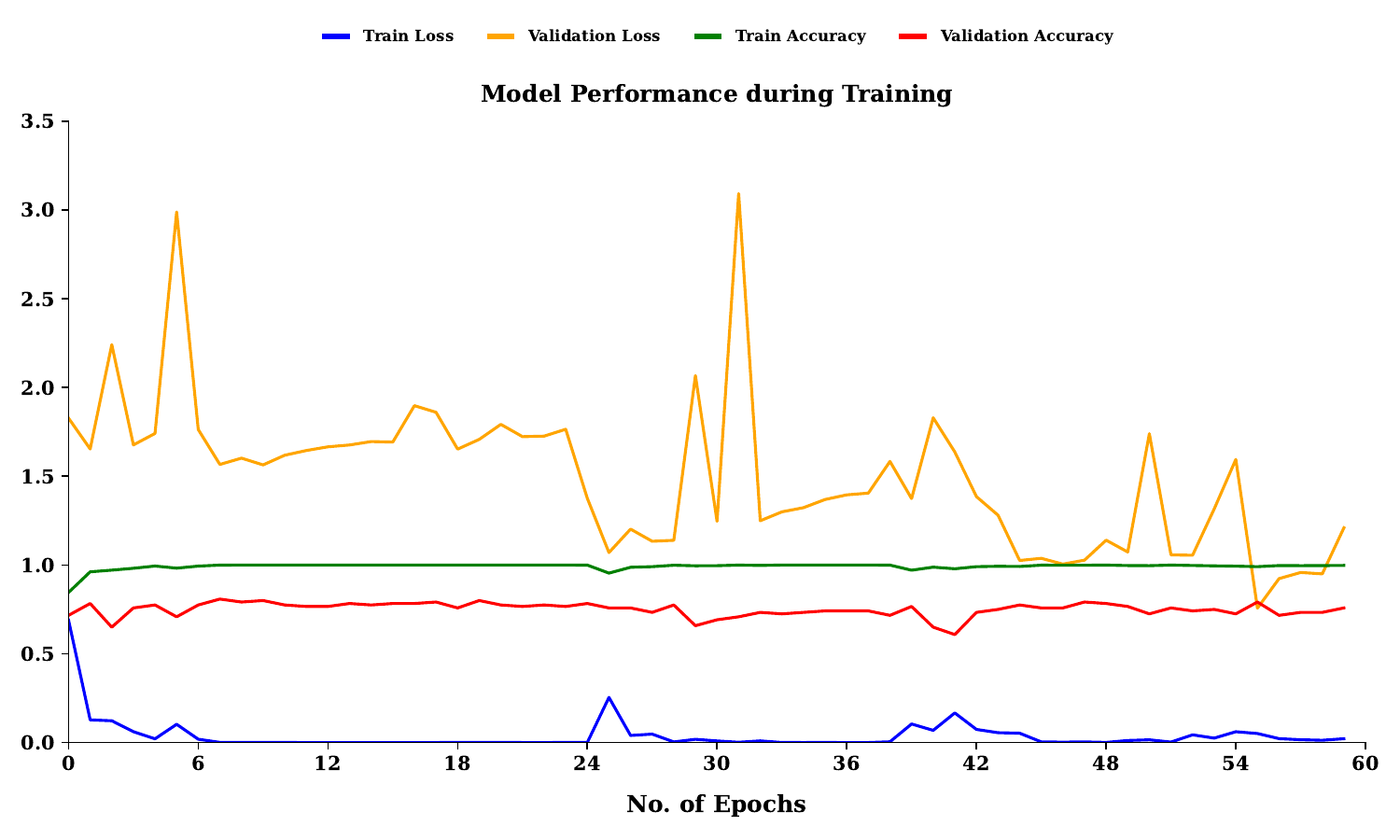}
        \caption{ResNet50}
        \label{fig:subAC7}
    \end{subfigure}
    ~
    \begin{subfigure}{0.32\textwidth}
        \includegraphics[width=\linewidth]{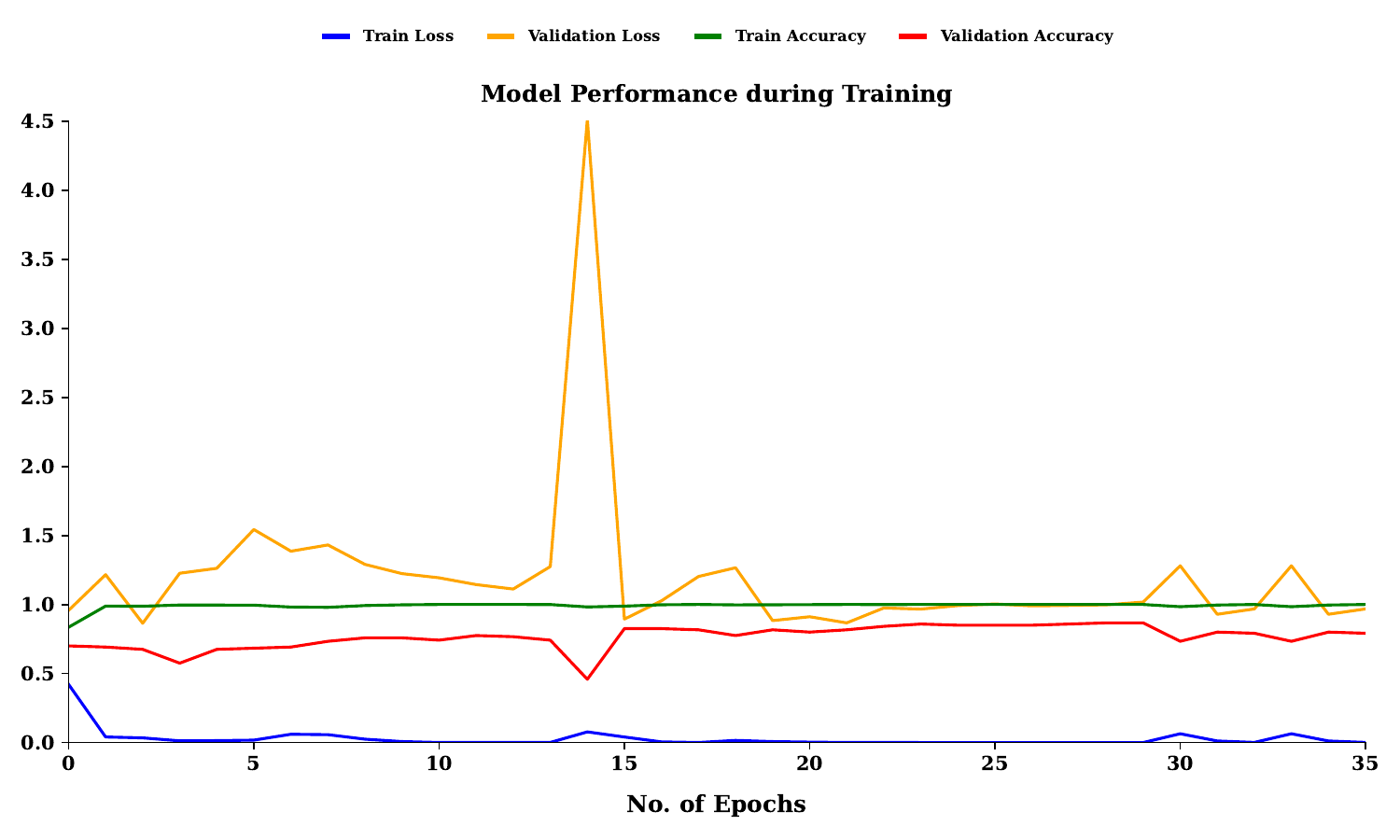}
        \caption{InceptionResNetV2}
        \label{fig:subAC8}
    \end{subfigure}
 
    \caption{Training and validation loss and accuracy across epochs for the eight pre-trained CNN backbones.}\label{fig:ACC}
\end{figure*}

Table \ref{tableClass} reports the eight pre-trained backbones on the FIVES test set, ordered by accuracy. ResNet101 is the strongest model, with an accuracy of 94.17\%, an F1 score of 0.9418, and the lowest log loss (0.2254); EfficientNetB0 is the weakest, at 88.33\%. The full ordering is ResNet101 $>$ DenseNet169 $>$ Xception $>$ InceptionV3 $>$ DenseNet121 $>$ InceptionResNetV2 $>$ ResNet50 $>$ EfficientNetB0. The confusion matrices are shown in Figure \ref{fig:CM} and the training and validation curves in Figure \ref{fig:ACC}. These results are analysed in Section \ref{findings}.


\begin{table*}[width=\textwidth,cols=7,pos=htbp]
\caption{Classification performance of the eight pre-trained CNN backbones on the FIVES test set, ordered by accuracy.}\label{tableClass}
\begin{tabular*}{\textwidth}{@{\extracolsep{\fill}}lcccccc@{}}
\toprule
\textbf{Model} & \textbf{Accuracy} & \textbf{Precision} & \textbf{Recall} & \textbf{F1 Score} & \textbf{Jaccard Score} & \textbf{Log Loss}\\
\midrule
ResNet101 & 0.9417 & 0.9435 & 0.9417 & 0.9418 & 0.8902 & 0.2254\\
DenseNet169 & 0.9333 & 0.9378 & 0.9333 & 0.9333 & 0.8751 & 0.9080\\
Xception & 0.9250 & 0.9284 & 0.9250 & 0.9252 & 0.8612 & 1.3931\\
InceptionV3 & 0.9167 & 0.9203 & 0.9167 & 0.9166 & 0.8480 & 0.8012\\
DenseNet121 & 0.9083 & 0.9092 & 0.9083 & 0.9075 & 0.8320 & 4.5509\\
InceptionResNetV2 & 0.9000 & 0.9049 & 0.9000 & 0.9008 & 0.8202 & 12.0282\\
ResNet50 & 0.8917 & 0.8948 & 0.8917 & 0.8926 & 0.8089 & 0.4883\\
EfficientNetB0 & 0.8833 & 0.8862 & 0.8833 & 0.8837 & 0.7947 & 0.6697\\
\bottomrule
\end{tabular*}
\end{table*}

\subsection{Retinal Blood Vessel Segmentation}
\begin{figure*}
    \centering
    \begin{subfigure}{1.0\textwidth}
    \centering
        \includegraphics[width=1.0\textwidth]{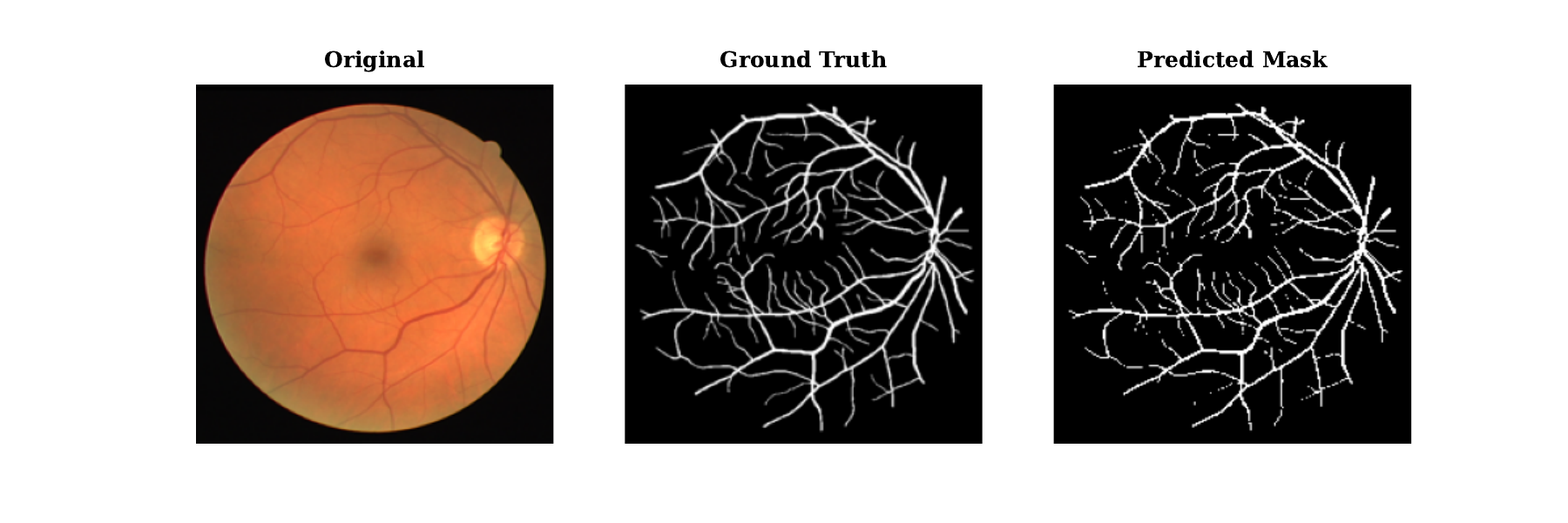}
        \caption{Attention U-Net: ResNet101V2}
        \label{fig:subSg1}
    \end{subfigure}
    \begin{subfigure}{1.0\textwidth}
    \centering
        \includegraphics[width=1.0\textwidth]{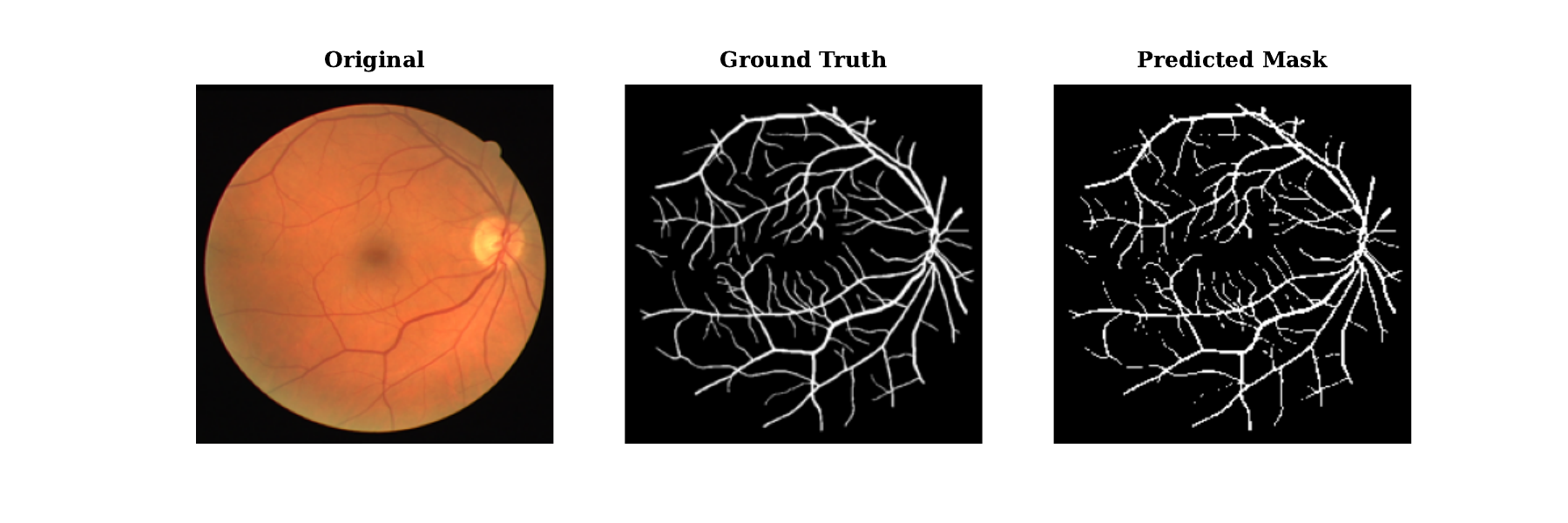}
        \caption{Attention U-Net: Resnet152V2}
        \label{fig:subSg2}
    \end{subfigure}
    \caption{Attention U-Net vessel segmentation on the DRIVE dataset with the two leading backbones.}\label{fig:driveseg}
\end{figure*}

\begin{figure*}
    \centering
    \begin{subfigure}{1.0\textwidth}
    \centering
        \includegraphics[width=1.0\textwidth]{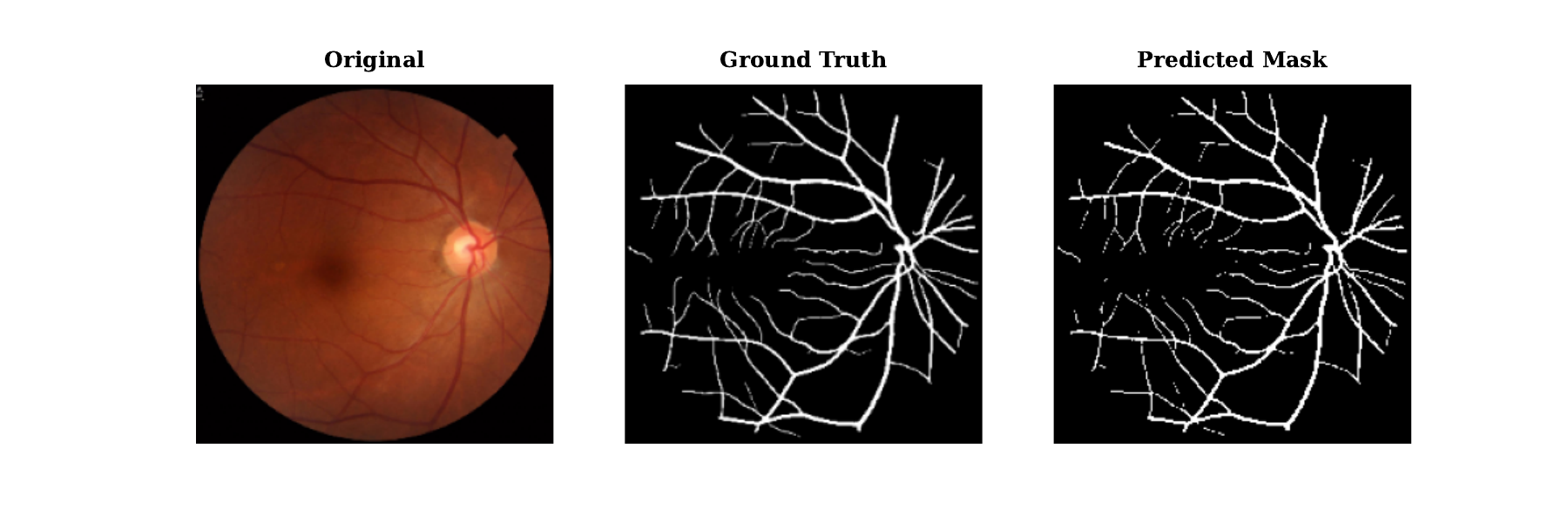}
        \caption{Attention U-Net: ResNet101V2}
        \label{fig:subFSg1}
    \end{subfigure}

    \begin{subfigure}{1.0\textwidth}
    \centering
        \includegraphics[width=1.0\textwidth]{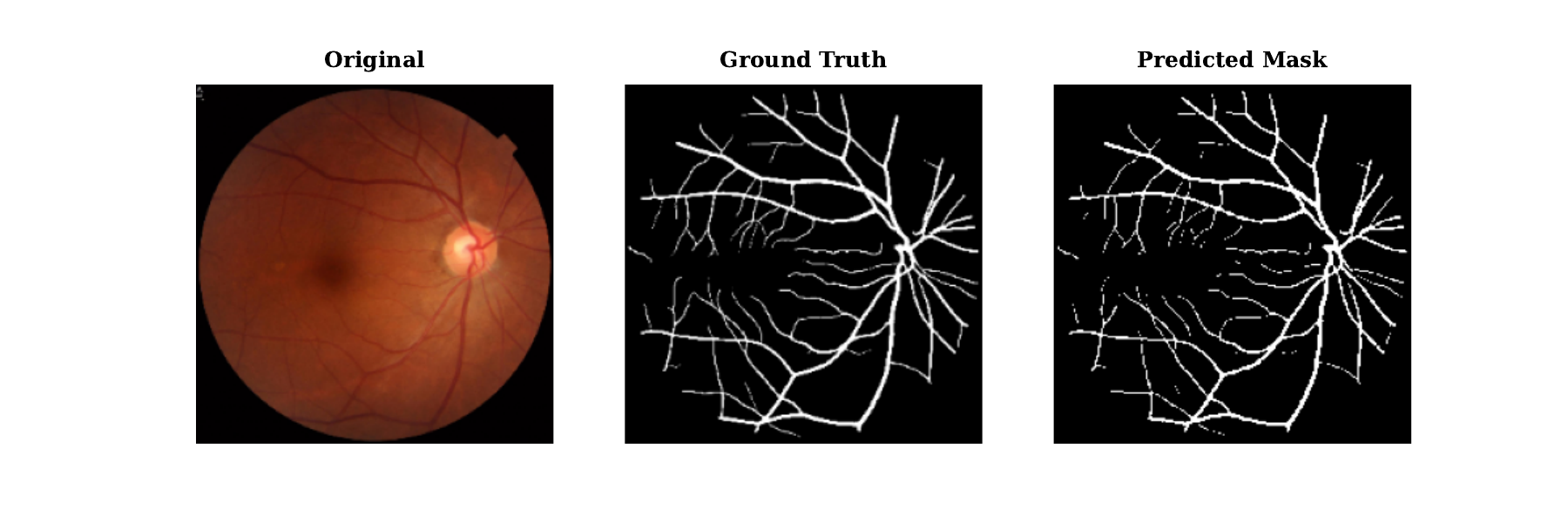}
        \caption{Attention U-Net: Resnet152V2}
        \label{fig:subFSg2}
    \end{subfigure}
    \caption{Attention U-Net vessel segmentation on the FIVES dataset with the two leading backbones.}\label{fig:Fiveseg}
\end{figure*}

Table \ref{ResDrive} and Table \ref{ResFives} report the ten U-Net variants on the DRIVE and FIVES test sets, using intersection over union (IoU), Dice coefficient, mean pixel accuracy (MPA), mean modified Hausdorff distance (MMHD), and mean surface Dice overlap (MSDO). On both datasets the architecture ordering is Attention U-Net $>$ TransUNET $>$ Swin-UNET, and within the Attention U-Net family the ResNet101V2 backbone is strongest, with IoU 0.6483 and Dice 0.7865 on DRIVE and IoU 0.7221 and Dice 0.8385 on FIVES. Qualitative segmentations for the two leading backbones are shown in Figure \ref{fig:driveseg} and Figure \ref{fig:Fiveseg}. These results are analysed in Section \ref{findings}.

\begin{table*}[width=\textwidth,cols=7,pos=htbp]
\caption{Segmentation performance of the ten U-Net variants on the DRIVE test set. Metrics are intersection over union (IoU), Dice coefficient, mean pixel accuracy (MPA), mean modified Hausdorff distance (MMHD), and mean surface Dice overlap (MSDO).}\label{ResDrive}
\begin{tabular*}{\textwidth}{@{\extracolsep{\fill}}llccccc@{}}
\toprule
\textbf{Model} & \textbf{Backbone} & \textbf{IoU} & \textbf{Dice} & \textbf{MPA} & \textbf{MMHD} & \textbf{MSDO}\\
\midrule
TransUNET & ResNet50V2 & 0.5273 & 0.6899 & 0.6470 & 3.852 & 0.0151\\
 & ResNet101V2 & 0.5169 & 0.6803 & 0.6400 & 4.0864 & 0.0188\\
 & ResNet152V2 & 0.5250 & 0.6879 & 0.6514 & 4.052 & 0.0167\\
\midrule
Attention U-Net & ResNet50V2 & 0.6460 & 0.7848 & 0.7953 & 2.7150 & 0.0070\\
 & ResNet101V2 & 0.6483 & 0.7865 & 0.8023 & 2.7341 & 0.0054\\
 & ResNet152V2 & 0.6474 & 0.7859 & 0.7859 & 2.6718 & 0.0069\\
 & DenseNet121 & 0.6427 & 0.7824 & 0.7747 & 2.7484 & 0.0090\\
 & DenseNet169 & 0.6444 & 0.7837 & 0.7769 & 2.6776 & 0.0072\\
 & DenseNet201 & 0.6422 & 0.7827 & 0.7749 & 2.6475 & 0.0074\\
\midrule
Swin-UNET & Swin Transformer & 0.4896 & 0.6569 & 0.6062 & 4.3675 & 0.0178\\
\bottomrule
\end{tabular*}
\end{table*}

\begin{table*}[width=\textwidth,cols=7,pos=htbp]
\caption{Segmentation performance of the ten U-Net variants on the FIVES test set. Metrics are as defined in Table \ref{ResDrive}.}\label{ResFives}
\begin{tabular*}{\textwidth}{@{\extracolsep{\fill}}llccccc@{}}
\toprule
\textbf{Model} & \textbf{Backbone} & \textbf{IoU} & \textbf{Dice} & \textbf{MPA} & \textbf{MMHD} & \textbf{MSDO}\\
\midrule
TransUNET & ResNet50V2 & 0.6436 & 0.7791 & 0.7758 & 3.7392 & 0.0312\\
 & ResNet101V2 & 0.6559 & 0.7898 & 0.7764 & 3.5491 & 0.0285\\
 & ResNet152V2 & 0.6522 & 0.7866 & 0.7696 & 3.5278 & 0.0319\\
\midrule
Attention U-Net & ResNet50V2 & 0.7175 & 0.8353 & 0.8512 & 2.8913 & 0.0201\\
 & ResNet101V2 & 0.7221 & 0.8385 & 0.8507 & 2.8009 & 0.0223\\
 & ResNet152V2 & 0.7199 & 0.8369 & 0.8331 & 2.8134 & 0.0378\\
 & DenseNet121 & 0.6872 & 0.8143 & 0.7866 & 3.2814 & 0.0591\\
 & DenseNet169 & 0.6718 & 0.8024 & 0.8115 & 3.5513 & 0.0235\\
 & DenseNet201 & 0.6468 & 0.7822 & 0.7587 & 3.6267 & 0.0293\\
\midrule
Swin-UNET & Swin Transformer & 0.5179 & 0.6765 & 0.5987 & 4.6090 & 0.0891\\
\bottomrule
\end{tabular*}
\end{table*}

\begin{figure*}
\centering
\includegraphics[width=0.7\textwidth]{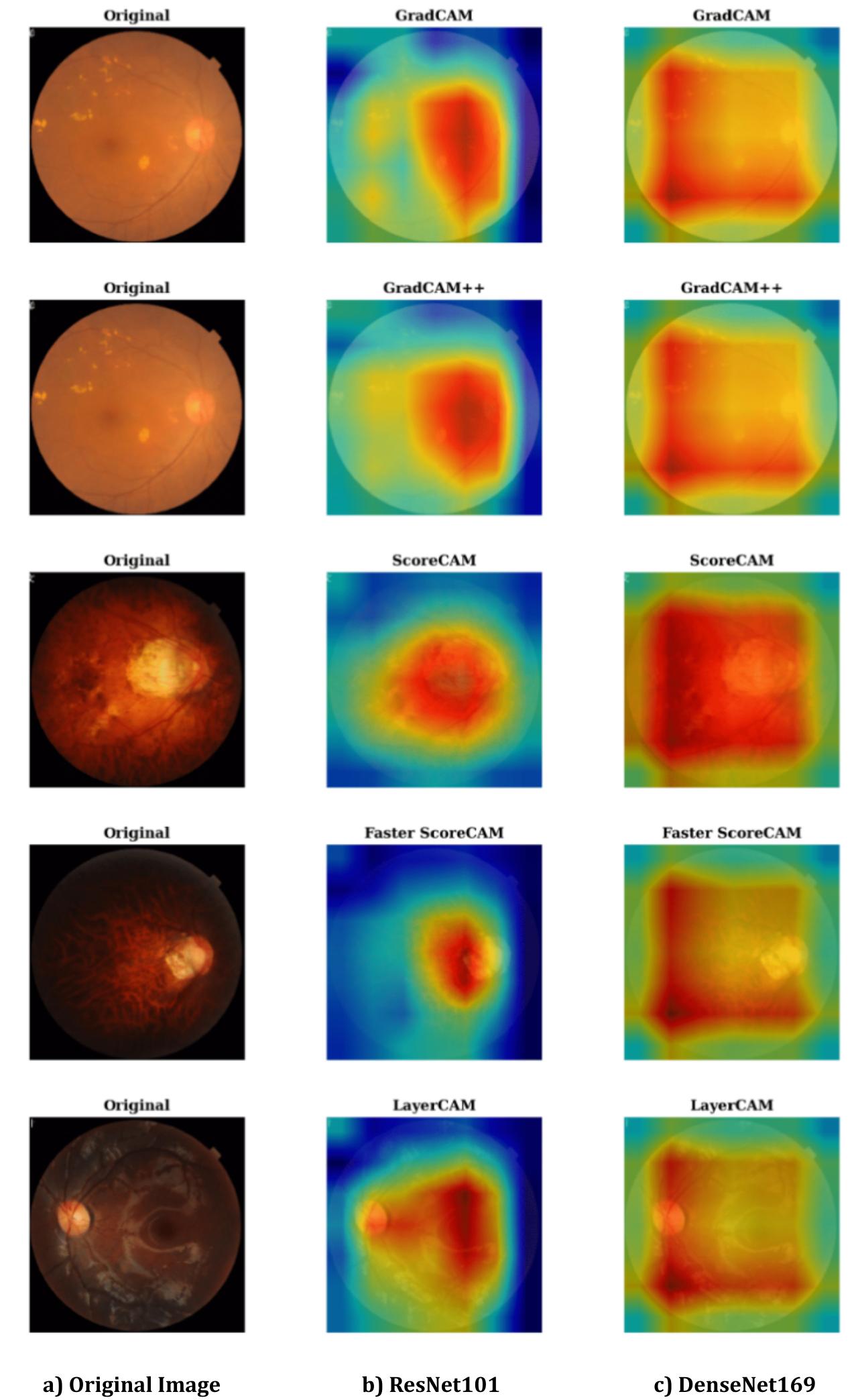}
\caption{The five gradient-based attribution methods applied to the same fundus image. Column (a) is the original image; columns (b) and (c) show the maps produced with the ResNet101 and DenseNet169 classifiers, respectively.}\label{fig:XAI}
\end{figure*}
\begin{table*}
\caption{Comparison of the proposed Attention U-Net with a ResNet101V2 backbone against previously reported results on the DRIVE and FIVES datasets.}\label{tab2}
\begin{tabular}{llllll}
\toprule
\textbf{Dataset} & \textbf{Study} & \textbf{Model} & \textbf{Metric} & \textbf{Reported} & \textbf{This work}\\
\midrule
\multirow{2}{*}{DRIVE} & Wang et al. \cite{Wang2021} & UNet++ & Dice coefficient & 78.63 & 78.65\\
 & Ko-Wei et al. \cite{RF8} & Custom U-Net & Intersection over union & 60.80 & 64.83\\
\midrule
\multirow{2}{*}{FIVES} & Ronneberger et al. \cite{Re} & U-Net & Dice coefficient & 81.74 & \multirow{2}{*}{83.85}\\
 & Yan et al. \cite{Intro7} & Custom U-Net & Dice coefficient & 81.65 & \\
\bottomrule
\end{tabular}
\end{table*}

\section{Discussion}\label{findings}
This section discusses the findings in relation to the four research questions introduced in Section \ref{methodology}, with emphasis on the implications of the results for retinal fundus classification, explainability, and clinical applicability. 
\cleardoublepage
\begin{tcolorbox}[
    enhanced,
    colback=rqbg,
    colframe=rqheader,
    colbacktitle=rqheader,
    coltitle=white,
    title=\textbf{RQ1: Backbone and family comparison for fundus classification},
    fonttitle=\bfseries,
    boxrule=0.5pt,
    arc=3pt,
    left=7pt, right=7pt, top=5pt, bottom=5pt,
    toptitle=1mm, bottomtitle=1mm
]
The eight backbones separate into three bands (Table \ref{tableClass}). The residual and densely connected families form the top band: ResNet101 (94.17\%) $>$ DenseNet169 (93.33\%). Xception and InceptionV3 form a middle band near 92\%. The remaining models fall below 91\%, with EfficientNetB0 last at 88.33\%, so the full spread is 5.84 points. Backbone depth within a family is the clearest single factor: ResNet101 $>$ ResNet50 by 5.00 points and DenseNet169 $>$ DenseNet121 by 2.50 points, in both cases the deeper variant winning. The compound-scaled EfficientNetB0, which trades depth for width and input scaling, is the weakest model, so on a four-class problem with a few hundred images per class raw representational depth is more useful than parameter efficiency. Calibration behaves differently from accuracy: InceptionResNetV2 and DenseNet121 reach log-loss values of 12.03 and 4.55 while classifying near 90\% correctly, which means their probability estimates are overconfident on the cases they get wrong; ResNet101 by contrast pairs the best accuracy with the lowest log loss (0.2254), so it is both the most accurate and the best-calibrated backbone. The confusion matrices in Figure \ref{fig:CM} and the loss curves in Figure \ref{fig:ACC} show that the leading models converge without severe over-fitting, whereas the lower-ranked models plateau earlier.
\end{tcolorbox}

\noindent For a four-class fundus classifier trained on FIVES, a deep residual backbone such as ResNet101 is the recommended choice: it gives the highest accuracy and, unlike several competitors, well-behaved class probabilities. Because accuracy and log loss do not agree across the table, a model intended for clinical triage should be selected on calibration as well as accuracy rather than on accuracy alone.

\begin{tcolorbox}[
    enhanced,
    colback=rqbg,
    colframe=rqheader,
    colbacktitle=rqheader,
    coltitle=white,
    title=\textbf{RQ2: Comparing gradient-based explanation methods and their dependence on the classifier},
    fonttitle=\bfseries,
    boxrule=0.5pt,
    arc=3pt,
    left=7pt, right=7pt, top=5pt, bottom=5pt,
    toptitle=1mm, bottomtitle=1mm
]
Figure \ref{fig:XAI} applies the five methods to the two strongest classifiers, ResNet101 and DenseNet169. Judged by how tightly the saliency concentrates on retinal structure relevant to the predicted class and how little falls on the background or the image border, the methods order as Score-CAM $>$ Grad-CAM++ $\approx$ Layer CAM $>$ Grad-CAM, with Faster Score-CAM close to Score-CAM but coarser at the boundaries. Score-CAM places high response on the vascular arcades and lesion regions and leaves the periphery mostly inactive; Grad-CAM and Layer CAM give broadly correct but more diffuse maps; Grad-CAM++ sharpens small structures at the cost of scattered activation. The quality of every method also tracks the classifier: with ResNet101 all five methods produce coherent, low-noise maps, while with DenseNet169 the same methods localise less precisely and fragment. The interpretability gap between the two backbones therefore mirrors their 0.84-point accuracy gap, which indicates that attribution quality is not a property of the method alone but of the method and the backbone together.
\end{tcolorbox}

\noindent The practical consequence is that the common default of reporting only Grad-CAM is not well founded, because the ranking of methods is not stable across architectures. On this data the pairing that best supports visual inspection by a clinician is Score-CAM computed on the strongest available backbone; a weaker backbone narrows the set of attribution methods that can be trusted.

\begin{tcolorbox}[
    enhanced,
    colback=rqbg,
    colframe=rqheader,
    colbacktitle=rqheader,
    coltitle=white,
    title=\textbf{RQ3: Attention and transformer U-Nets for vessel segmentation, and the role of the backbone},
    fonttitle=\bfseries,
    boxrule=0.5pt,
    arc=3pt,
    left=7pt, right=7pt, top=5pt, bottom=5pt,
    toptitle=1mm, bottomtitle=1mm
]
The decoder mechanism matters more than the backbone. On both datasets the architecture ordering is Attention U-Net $>$ TransUNET $>$ Swin-UNET (Table \ref{ResDrive}, Table \ref{ResFives}). On DRIVE the gap is large: every Attention U-Net configuration exceeds IoU 0.64 while Swin-UNET reaches only 0.4896, a difference of more than 0.15 IoU. The attention gates on the skip connections, which suppress background responses, are therefore more useful for thin, low-contrast vessels than either the hybrid convolutional-transformer bottleneck of TransUNET or the fully transformer encoder of Swin-UNET; the pure transformer design is the least suited to this task at the available data scale. Within the Attention U-Net family the ResNet101V2 backbone is best on both datasets (DRIVE IoU 0.6483, Dice 0.7865; FIVES IoU 0.7221, Dice 0.8385), and the residual backbones as a group give lower mean modified Hausdorff distances, meaning cleaner vessel boundaries. On DRIVE the six backbones lie within a 0.006 IoU band, whereas on the higher-resolution FIVES images the residual backbones pull clear of the dense ones (DenseNet201 $\rightarrow$ IoU 0.6468). Against prior work (Table \ref{tab2}), the proposed Attention U-Net with ResNet101V2 improves DRIVE IoU from 60.80 to 64.83 over a custom U-Net and reaches a DRIVE Dice of 78.65, comparable to a previously reported UNet++ result of 78.63; on FIVES its Dice of 83.85 is above the two reported U-Net baselines of 81.74 and 81.65.
\end{tcolorbox}

\noindent Attention U-Net with a ResNet101V2 backbone is the configuration we recommend for retinal vessel segmentation: it is the strongest on both datasets and improves clearly on a plain U-Net in region overlap, while its Dice score is on par with a stronger UNet++ baseline. Transformer components help only when they are combined with a convolutional stream; a transformer-only encoder is not competitive here.

\begin{tcolorbox}[
    enhanced,
    colback=rqbg,
    colframe=rqheader,
    colbacktitle=rqheader,
    coltitle=white,
    title=\textbf{RQ4: Whether classification design choices transfer to segmentation},
    fonttitle=\bfseries,
    boxrule=0.5pt,
    arc=3pt,
    left=7pt, right=7pt, top=5pt, bottom=5pt,
    toptitle=1mm, bottomtitle=1mm
]
One choice transfers and one does not. Backbone depth within the residual family transfers cleanly: ResNet101 is the best classifier and the ResNet101V2 backbone is the best Attention U-Net encoder, and in both pipelines the deeper member of a family beats the shallower one. The densely connected backbones are competitive for classification but fall behind the residual backbones for segmentation on the higher-resolution FIVES images, so segmentation is the more backbone-sensitive of the two tasks. The value of transformer encoders does not transfer: the best classifier needs no transformer component, and the fully transformer Swin-UNET is the weakest segmentation model on both datasets ($0.4896 \le$ IoU $\le 0.5179$), well below every convolutional or hybrid variant. Read together with RQ2, where a stronger backbone widened the set of reliable attribution methods, the three findings point in the same direction: a deep residual backbone benefits accuracy, boundary quality, and explanation quality at once.
\end{tcolorbox}

\noindent For retinal fundus analysis at this data scale we therefore recommend a single design family across both tasks: a deep residual backbone, an attention-gated U-Net for segmentation, and Score-CAM for explanation, in preference to a transformer-first design. Backbone choice can be treated as a shared decision for the two pipelines, whereas the decoder and the explanation method must be chosen per task.

\section{Limitations}\label{limitation}
Several limitations should be kept in mind when interpreting the results.
(i) The evaluation rests on a single training and test split for each task, without cross-validation or repeated runs, so the reported metrics carry no variance or significance estimate and small differences between models should be read with caution.
(ii) The datasets are modest in size, with 800 images in FIVES and 40 in DRIVE, and were acquired from a limited set of populations and cameras; generalisation to other imaging devices, ethnicities, and acquisition protocols is therefore untested.
(iii) The classification problem uses the four balanced FIVES categories, whereas real screening data are strongly imbalanced and contain comorbid and out-of-distribution cases that are not represented here.
(iv) The classification and segmentation pipelines were developed and tuned separately rather than jointly optimised, so the cross-task observations in Section \ref{findings} reflect a comparison of two independent studies and not a controlled experiment in which one factor is varied at a time.
(v) The comparison of explanation methods is qualitative and limited to the two strongest backbones; no quantitative faithfulness metric was computed and no reader study with ophthalmologists was carried out, so the ranking of methods reflects visual inspection rather than measured localisation accuracy.
(vi) The mean surface Dice overlap values were internally inconsistent and were excluded from the analysis, which leaves the boundary-level assessment of the segmentation models incomplete.
(vii) The per-model epoch schedules were set manually rather than by a systematic search.
(viii) The segmentation loss and decision threshold follow common practice rather than task-specific tuning, so the reported numbers are lower bounds on what each architecture can achieve.
(ix) The comparison with prior work in Table \ref{tab2} is drawn from figures reported in heterogeneous experimental settings; the baselines were not re-run here, so the gaps should be read as indicative rather than as head-to-head differences.

\section{Future Work}\label{future}
The single most important next step is an external, multi-dataset validation with proper uncertainty reporting; the remaining directions address the interpretability, data, and deployment gaps identified above.
Firstly, the experiments should be repeated with k-fold cross-validation and multiple random seeds, reporting confidence intervals and significance tests, and validated on further public datasets such as STARE, CHASE\_DB1, HRF, Messidor, and EyePACS; we expect the margin between the residual and the transformer variants to narrow as the training set grows, and confirming or refuting this would clarify how far the present findings are data-scale artefacts.
Secondly, the interpretability analysis should be made quantitative through deletion and insertion curves, pointing-game accuracy, and model-randomisation sanity checks, and complemented by a reader study in which ophthalmologists rate the clinical usefulness of each saliency map, so that the qualitative advantage observed for Score-CAM can be tested against a measurable criterion.
Thirdly, the classification setup should be extended to imbalanced, multi-label, and open-set conditions and combined with clinical metadata such as patient age and history, which are known to carry diagnostic signal beyond the image.
Fourthly, on the segmentation side, task-specific loss functions, a corrected surface-based metric, and semi-supervised or multi-modal fusion with optical coherence tomography angiography should be explored to handle data scarcity and vessels that are occluded by pathology.
Finally, probability calibration methods such as temperature scaling, together with lightweight architecture variants, should be studied so that the predicted probabilities and the inference speed meet the requirements of real-time clinical screening.

\section{Conclusion}\label{conclusion}
This work studied two pipelines for retinal fundus analysis on public data: disease classification with eight pre-trained convolutional backbones together with five gradient-based explanation methods, and retinal blood vessel segmentation with ten U-Net variants spanning attention-gated, hybrid, and fully transformer designs. In classification, the backbone ranking was ResNet101 (94.17\%, F1 0.9418, log loss 0.2254) $>$ DenseNet169 $>$ Xception $>$ $\cdots$ $>$ EfficientNetB0 (88.33\%), and within every family the deeper variant won (ResNet101 $>$ ResNet50, DenseNet169 $>$ DenseNet121). Among the explanation methods the ordering was Score-CAM $>$ Grad-CAM++ $\approx$ Layer CAM $>$ Grad-CAM, and attribution quality rose and fell with backbone accuracy, so the choice of method and the choice of network are coupled rather than independent. For segmentation, Attention U-Net $\succ$ TransUNet $\succ$ Swin-UNet held on both DRIVE and FIVES, with Attention U-Net~$\oplus$~ResNet101V2 the best configuration overall: IoU 0.6483 / Dice 0.7865 on DRIVE and IoU 0.7221 / Dice 0.8385 on FIVES, which lifts a previously reported custom U-Net from IoU $60.80 \rightarrow 64.83$ and reaches a Dice score $\approx$ that of a stronger UNet++ baseline. Reading the two pipelines together, a deep residual backbone improved classification accuracy, segmentation boundary quality, and explanation quality at the same time ($\uparrow$ on all three axes), whereas a transformer-only encoder did not transfer across tasks (Swin-UNet last in both). We therefore recommend, for retinal fundus analysis at this data scale, a deep residual backbone $+$ an attention-gated U-Net for segmentation $+$ Score-CAM for interpretation. The benchmarks and the cross-task analysis reported here give a reference point for future work, whose priorities are external multi-dataset validation with formal uncertainty reporting and a quantitative, clinician-in-the-loop evaluation of interpretability.

\section*{Authors’ contributions}
\textbf{Fatema Tuj Johora Faria:} Conceptualization, Methodology, Investigation, Software, Writing - Original Draft. \textbf{Mukaffi Bin Moin:} Conceptualization, Methodology, Investigation, Software, Writing - Original Draft. \textbf{Pronay Debnath:} Validation,  Software, Writing - Review \& Editing. \textbf{Asif Iftekher Fahim:} Validation,  Software, Writing - Review \& Editing. \textbf{Faisal Muhammad Shah:} Formal analysis, Writing - Review \& Editing, Supervision.
\section*{Declarations}

\subsection*{Ethical Approval and Consent to participate}
Not Applicable.

\subsection*{Human and Animal Ethics}
Not Applicable.

\subsection*{Conﬂicts of Interest}
The authors declare that they have no conflicts of interest.

\subsection*{Funding}
This research was carried out with no external funding.

\subsection*{Code Availability}The code and supporting files for this study, publicly accessible for reference and use, can be found on at: 
\href{https://github.com/fatemafaria142/Retinal-Fundus-Classification-using-XAI-and-Segmentation}{GitHub}\footnote{\href{https://github.com/fatemafaria142/Retinal-Fundus-Classification-using-XAI-and-Segmentation}{https://github.com/fatemafaria142/Retinal-Fundus-Classification-using-XAI-and-Segmentation}}




\bibliography{Reference}

\end{document}